\documentclass[11pt]{article}
\pdfoutput=1
\usepackage{epsfig}
\usepackage{amsfonts}
\usepackage{amscd}
\usepackage{latexsym}
\usepackage{amsmath,amssymb}
\usepackage{verbatim}
\usepackage{mathrsfs}
\usepackage{setspace}
\usepackage{color}
\usepackage{fancyhdr}
\usepackage{cite}
\usepackage[hidelinks]{hyperref}
\usepackage{dsfont}
\usepackage{tikz}
\usepackage{slashed}
\usepackage{draft} 
\usepackage{hyperref}
\usepackage{graphicx,color,subfig}
\usepackage{cite}
\usepackage{draft} 
\usepackage{graphicx,color,subfig}
\usepackage{cite}
\usepackage{comment}
\usepackage{array}
\usepackage{tabularx}
\usepackage{mciteplus}
\usepackage{skak}
\DeclareFontFamily{OT1}{pzc}{}
\DeclareFontShape{OT1}{pzc}{m}{it}{<-> s * [1.10] pzcmi7t}{}
\DeclareMathAlphabet{\mathpzc}{OT1}{pzc}{m}{it}

\begin{document}

\begin{titlepage}

\begin{center}

\hfill \\
\hfill \\
\vskip 1cm

\title{\Large \bf{Dual resonant amplitudes from Drinfel'd twists}}

\author{Rishabh Bhardwaj and Shounak De}

\address{Department of Physics, Brown University,
	182 Hope Street, \\Providence, RI 02912, U.S.A.}

\vspace{1.0cm}

{{\tt rishabh\_bhardwaj@brown.edu \& shounak\_de@brown.edu}}

\end{center}

\vspace{1.0cm}

\begin{abstract}
We postulate the existence of a family of dual resonant, four-point tachyon amplitudes derived using invertible coproduct maps called Drinfel'd twists. A sub-family of these amplitudes exhibits well-defined ultraviolet behaviour, namely in the fixed angle high-energy and Regge scattering regimes. This discovery emerges from a systematic study of the set of observables that can be constructed out of $q$-deformed worldsheet CFTs with the underlying conformal group being the quantum group $SU(1,1)_q$. We conclude our analysis by discussing the possibility (or the lack thereof) of known $q$-deformations of the Veneziano amplitude as an observable in such theories, in particular, the Coon amplitude.

\end{abstract}

\vfill

\end{titlepage}

\eject
\tableofcontents

\section{Introduction}
\label{sec:intro}

The notion of \textit{quantum groups} first appeared in physics during the 1980s in the context of quantum integrable systems  \cite{Kulish1983QuantumLP,Faddeev:1987ih,Witten:1989rw}. Consequently, mathematicians like Drinfeld and Jimbo formalized quantum groups as a particular class of Hopf algebra and which deform enveloping Lie algebras \cite{drinfel'd1985hopf, Jimbo:1985zk}. Such new algebraic structures arose particularly in the
framework of the quantum inverse scattering method \cite{Sklyanin:1991ss}, and the quantum Yang-Baxter equations became a unifying basis of all such investigations. Over the past four decades, the theory of quantum groups has developed rapidly with widespread applications to conformal field theories (CFTs) \cite{Bernard:1989jq,Alvarez-Gaume:1988izd, ALVAREZGAUME1989155}, knot and link invariants \cite{Takhtajan1990}, and even quantum gravity \cite{Majid:1994cy, Jevicki:1998rr}. We refer the interested reader to \cite{lusztig2010introduction, majid_1995, Takhtajan1990,kassel2012quantum,Castellani:1996se} for excellent reviews on the subject.

In 1989, Drinfel'd introduced a seminal procedure for twisting such Hopf algebras \cite{drinfel'd1989quasi,drinfeld1990quasi} by showing that there exists an isomorphism between quantum symmetries encoded in the Hopf algebra and the corresponding universal covering group of any given (semi-simple) Lie algebra. These invertible elements of Hopf algebras have been coined \textit{Drinfel'd twists} and are widely applicable in the field of quantum groups. In physics, the Drinfel'd twist has been useful from the perspective of integrability, particularly in the context of solving equations that arise in the algebraic Bethe ansatz \cite{Maillet:1996yy}, as well as in the context of the AdS/CFT correspondence \cite{Beisert:2005if,vanTongeren:2015uha}. More recently, it has also found applications in the field of non-commutative gauge theories \cite{vanTongeren:2015uha,Meier:2023lku}.

The application of quantum groups to CFTs has been rather prominent. In particular, there have been several approaches to construct $q$-deformed worldsheet CFTs based on $q$-deformations of the underlying conformal group, thereby taking one to the realm of quantum groups. Some early works \cite{Bernard:1989jq,Matsuzaki:1991vu} in this regard include the construction of $q$-CFTs by exploiting the Hopf algebra structure of the quantum group $SU(1,1)_q$ to furnish $q$-deformed correlation functions as solutions to $q$-Ward identities. This analysis leads to $q$-deformed amplitudes by utilising conventional gauge-fixing conditions of string theory, which serve as $q$-deformations to the well-known Veneziano amplitude. Other such attempts include the construction of an operator formulation of a $q$-deformed dual resonant amplitude using an infinite set of $q$-harmonic oscillators \cite{Chaichian:1992hr} and the study of the $q$-deformation of the full Virasoro algebra \cite{Chaichian:1990rt,chaichian1990q}. The analysis in our paper will give a detailed review of some of these works and shed new light on some old results. 

Apart from such $q$-deformed dual resonant amplitudes derived from the perspective of quantum groups, several orthogonal investigations have postulated the existence of well-known Veneziano deformations.\footnote{More recent investigations into the uniqueness of the Veneziano amplitude have revealed multi-parameter spaces of four-point amplitudes satisfying the bootstrap constraints \cite{Cheung:2022mkw,Cheung:2023adk,Cheung:2023uwn,Geiser:2022exp}. However, the analysis in our paper will solely be restricted to one-parameter deformations.} In particular, the Coon amplitude \cite{Coon:1969yw} serves as a well-known $q$-deformed amplitude, which interpolates between the Veneziano amplitude as $q \to 1$ and a scalar field theory amplitude as $q \to 0$. Following its inception, many aspects of the Coon amplitude have been studied in some detail, including its $n$-point generalization \cite{Baker:1970vxk}, its operator formulations \cite{Arik:1973eb,Coon:1972te,Chaichian:1992hr} and constraints due to unitarity \cite{Figueroa:2022onw, Chakravarty:2022vrp,Bhardwaj:2022lbz,Jepsen:2023sia}. Despite attempts to understand the physical origins of the Coon amplitude \cite{Maldacena:2022ckr}, a worldsheet realization still remains elusive. While the utilisation of the quantum group $SU(1,1)_q$ \cite{Bernard:1989jq,Romans:1988qs,Romans:1989di} provided a necessary first step towards this direction, there have been recent attempts from a vertex operator formulation \cite{Li:2023hce}.

Our work attempts to initiate a systematic investigation into the set of observables emanating from $q$-CFTs whose underlying conformal group is the quantum group $SU(1,1)_q$. We employ a two-pronged approach to tackle the above problem: i) First, we review existing literature on the construction of $q$-CFTs using the Hopf algebra structure of the quantum group $SU(1,1)_q$ initiated in \cite{Bernard:1989jq,Matsuzaki:1991vu} before extending such constructions to well-known representations of its Quantum Universal Enveloping algebras. ii) Second, we elucidate for the first time, the construction of a $q$-deformed CFT using the Drinfel'd twist corresponding to $SU(1,1)_q$. Our analysis allows for the construction of $q$-deformed amplitudes by employing open string gauge-fixing conditions of bosonic string theory on $q$-deformed correlation functions. A significant byproduct of constructing field theories out of these Drinfel'd twists is the existence of a class of $q$-deformed amplitudes that are \textit{dual resonant}. We also discuss the potential realisation (or the lack thereof) of the Coon amplitude in the worldsheet $q$-CFTs constructed in our work.

This paper is structured as follows. In Section \ref{sec:setup}, we review existing constructions of $q$-deformed worldsheet CFTs before extending this formalism to well-known representations of the quantum universal enveloping group $\mathcal{U}_q(SU(1,1))$. In Section \ref{sec:qCFTDrinfel'd}, we present a formalism to construct a worldsheet $q$-CFT using twisted Ward identities arising out of Drinfel'd twists. On appropriate analytic continuation, the $q$-deformed amplitudes constructed out of these twists exhibit dual resonance and well-defined asymptotic behaviour, which is discussed in Section \ref{sec:asympproptwistedamp}. We discuss the possibility of a worldsheet realisation of the Coon amplitude using such $q$-CFT constructions in Section \ref{sec:Coonamp}, before concluding our analysis with discussions and potential avenues for future work in Section \ref{sec:discussions}.

\section{\texorpdfstring{$q$-CFTs from non-invertible Hopf algebras}{q-CFTs from non-invertible Hopf algebras}} \label{sec:setup}
In this section, we begin the first stage of our analysis where we construct $q$-CFTs using Hopf algebra maps that are \textit{non-invertible} when one flows away from $q=1$ (where the undeformed worldsheet CFT of string theory lives). We begin by reviewing existing literature on this subject, in particular, the construction of a $q$-CFT arising out of a well-known $q$-deformation of the $SU(1,1)$ conformal group put forth by Drinfel'd and Jimbo \cite{drinfel'd1985hopf,Jimbo:1985zk} and independently by Kulish and Reshetikhin \cite{Kulish1983QuantumLP}. 
We provide two more explicit $q$-CFT constructions out of such non-invertible Hopf algebra maps in subsections \ref{sec:MS_deformation} and \ref{sec:woronowiczdeformation}. We conclude the section in \ref{sec:spaceofQUE-algebras} by establishing functional isomorphisms between a whole host of such deformations and pinpoint a common feature in the observables of such deformed worldsheet theories. We kindly refer the reader to appendix \ref{app:quantum_group} as and when necessary for mathematical preliminaries and terminologies used in the rest of this section.  

\subsection{Revisiting the Drinfel'd-Jimbo deformation}\label{sec: DJ_deformation}
The analysis of \cite{Bernard:1989jq} uses a $q$-deformed universal enveloping algebra $\mathcal{U}_q(SU(1,1))$ defined by the following Lie algebra structure:
\begin{align}
\left[\mathcal{L}_{+},\mathcal{L}_{-}\right] = -\frac{\mathcal{H}^2-\mathcal{H}^{-2}}{q-q^{-1}}~~,\quad q^{\pm 1} \mathcal{L}_{\pm} \mathcal{H} = \mathcal{H} 
 \mathcal{L}_{\pm}~\label{eq:DJ_algebra},    
\end{align}
where the usual $SU(1,1)$ relations are recovered in the $q \to 1$ limit under the identification $\mathcal{H} = q^{\mathcal{L}_0}$. It is equipped with a Hopf algebra structure $\mathcal{U}_q \rightarrow \mathcal{U}_q \hspace{0.3mm}\otimes \hspace{0.3mm}\mathcal{U}_q$ defined by the $q$-deformed comultiplication (or the coproduct) map $\Delta_q$:
\begin{align}\label{eq:hopf_algebra_of_DJ}
\Delta_q(\mathcal{L}_{\pm}) = \mathcal{L}_{\pm} \otimes \mathcal{H} + \mathcal{H}^{-1} \otimes \mathcal{L}_{\pm}~, \quad \Delta_q(\mathcal{H}) = \mathcal{H} \otimes \mathcal{H}~.
\end{align}
There exists a representation of $\mathcal{U}_q(SU(1,1))$ as operators on the space of functions $f(z)$ with conformal weight $\Delta$:
\begin{align}\label{eq:representation_sl2c_generators}
\mathcal{L}_{-} f(z) = \frac{1}{z} \frac{f(zq)-f(zq^{-1})}{q-q^{-1}}~,~ \mathcal{L}_{+} f(z) = z \frac{q^{2\Delta}f(zq)-q^{-2\Delta}f(zq^{-1})}{q-q^{-1}}~,~ \mathcal{H}f(z) = q^{\Delta} f(zq)~.
\end{align}
The authors of \cite{Bernard:1989jq} begin with the following ansatz for the $N$-point correlator
\begin{align}
    \left\la \prod_{i=1}^N \phi_i(z_i)\right\ra_q = \prod_{i<j}^{N} z_{i}^{\gamma_{ij}} G_q \left(x_{ij}|~a_{ij}; \gamma_{ij}\right)~,\label{eq:N-p_ansatz_BC}
\end{align}
where $\phi_i(z_i)$ is a conformal scalar primary of dimension $\Delta_i$\footnote{One can define $\phi_i(z_i)$ as linear tachyonic vertex operators of the worldsheet embedding field $X^{\mu}(z_i)$ defined using $q$-deformed creation and annihilation operators; the interested reader is referred to \cite{Bernard:1989jq},\cite{Awata:1996xt} for more details.} and $x_{ij} \equiv z_j/z_i$. The coefficients $a_{ij}, \gamma_{ij}$ are to be determined by requiring that the ansatz solves the $N$-point $q$-Ward identities, as in conventional, undeformed CFTs. Subsequently, we work with $q$-deformed correlators unless otherwise specified and hence drop the $q$ subscript from the correlators. 
We note that these $N$-point functions are not completely fixed by the underlying conformal symmetries for $N \geq 4$. In particular, at the level of $N=4$ it is well known that the full form of the correlation function is fixed up to a function $F(u,v)$ of the conformal cross-ratios $u~\text{and}~v$ \cite{DiFrancesco:1997nk}. In the following analysis, we shall work with the four-point ansatz whose conformally invariant function is set to unity, i.e., $F(u,v)$=1. This is also the case for the four-point worldsheet correlators involving tachyonic vertex operators in string theory. Since unity is a fixed point of the deformation (i.e., $[1]_q=1$), this is a well-motivated choice for the corresponding $q$-CFT four-point correlator as well.\footnote{We will stick to this choice for the conformally invariant function $F(u,v)$ in the subsequent analysis, until it becomes relevant to our discussion again in section \ref{sec:Coonamp}.} The kernel $G_q \left(x_{ij}|~a_{ij}; \gamma_{ij}\right)$ in the ansatz (\ref{eq:N-p_ansatz_BC}) is defined as the following infinite product representation 
\begin{equation}
    G_q\left(x|~a; \gamma\right) = \frac{\prod_{n=0}^{\infty}(1-xq^{2n+a})}{\prod_{n=0}^{\infty}(1-xq^{2n+a+2\gamma})}~,
    \label{eq:DJ_ansatz_seed}
\end{equation}
and reduces to $z_i^{\gamma}G_q\left(x_{ij}|~a; \gamma\right) \overset{q \to 1}{\longrightarrow} z_i^{\gamma}(1-x_{ij})^{\gamma} = (z_i-z_j)^{\gamma}$ in the classical limit, as required by the standard four-point CFT functions. It also satisfies the following $q$-difference equation
\begin{align}
(1-xq^a) G_q \left(x q^2|~a; \gamma\right) = (1-xa^{a+2\gamma}) G_q\left(x|~a; \gamma\right)~\label{eq:Gq_identity},    
\end{align}
which will be used throughout to solve the $q$-deformed Ward identities. An important ingredient to construct $N$-point $q$-Ward identities is the $(N-1)$-fold coproduct defined as 
\begin{align}
\Delta^{(N-1)}_q(X)= \overbrace{(\Delta_q\otimes\text{id}\otimes\dots\otimes \text{id})}^{N-2}\cdot \overbrace{(\Delta_q\otimes\text{id}\otimes\dots\otimes \text{id})}^{N-3}\dots(\Delta_q\otimes\text{id})\cdot\Delta_q(X)~.
\label{eq:N-foldcoproduct}
\end{align}
Using the above $(N-1)$-fold coproduct, the $q$-Ward identity of $\mathcal{H}_0$ is given by \cite{Jimbo:1985zk}
\begin{align}
    \Delta_q^{(N-1)}(\mathcal{H})\la\phi_1(z_1)\dots\phi_N(z_N)\ra &= q^{\sum_{i=1}^N \Delta_i}\la\phi_1(qz_1)\dots\phi_N(qz_N)\ra \nonumber\\
    &=q^{\sum_{i=1}^N \Delta_i+\sum_{i<j} \gamma_{ij}}\la\phi_1(z_1)\dots\phi_N(z_N)\ra~ \nonumber\\
    &=\la\phi_1(z_1)\dots\phi_N(z_N)\ra~,
\end{align}
which leads to the following relationship between the conformal dimensions $\Delta_i$ and $\gamma_{ij}$
\begin{align}
\sum_{i<j} \gamma_{ij} = -\sum_{i=1}^N \Delta_i~.
\label{eq:H0_Ward_identity}
\end{align}
The $q$-Ward identity for the other two $SU(1,1)_q$ generators using (\ref{eq:N-foldcoproduct}) is derived to be  
\begin{align}
    \Delta_q^{(N-1)}(\mathcal{L}_{n}) &= \sum_{i=1}^{N}q^{-\sum_{j=1}^{i-1}\Delta_j}\la\phi_1(q^{-1}z_1)\dots \phi_{i-1}(q^{-1}z_{i-1})\hat{\phi}_{i}(z_i)\phi_{i+1}(qz_{i+1})\dots\phi_N(qz_N)\ra~, 
\label{ActionofL_NBC}
\end{align}
where we have defined $\hat{\phi}(z) = \mathcal{L}_n\phi(z)$. Making the simplifying assumption that $\Delta_i=\Delta$ for any $i$, using the Ward identities for the $\mathcal{L}_{\pm}$ generators on the four-point correlator and plugging in the ansatz (\ref{eq:N-p_ansatz_BC}), we observe that we have the following constraints over the unknowns:
\begin{align}
a_{i,i+1} &= 2\Delta~~, \qquad \forall i\in\{1,2,3\}  \nonumber \\
a_{14}+2\gamma_{14} &= -2\Delta~,~~~\gamma_{13} = 0 =\gamma_{24}~, \label{eq:Lm_constraints_BC1}
\end{align}
along with 
\begin{align}
    \sum_{i\neq j}\gamma_{ij} = -2\Delta~,~~~~~~~\forall i\in\{1,2,3,4\}
\end{align}
which is indeed the result stated in \cite{Bernard:1989jq}. To construct the corresponding $q$-string amplitude, we first need an identification that parameterises the unknowns in the correlator (\ref{eq:N-p_ansatz_BC}) with the amplitude invariants. As considered in \cite{Bernard:1989jq}, we identify $\gamma_{ij}= \alpha_i\cdot\alpha_j$ along with $\sum_i\alpha_i=0$ and $\alpha_i^2 = 2\Delta$. We further define the Mandelstam variables as $s \equiv -(\alpha_1+\alpha_2)^2$ along with $t \equiv -(\alpha_2+\alpha_3)^2$ and set $\Delta$ to be unity.\footnote{We work in the $(+,-,-,-)$ signature for which the sign of the Mandelstam variables are  $s<0,t>0,u>0$.} In order to define the amplitude, we consider integrating over the natural Haar measure on the $q$-deformed sphere $S_q^2$, the group manifold of $SU(1,1)_q$. Further, one uses the ordinary notion of gauge fixing as in the case of open strings, i.e. $\{z_1=0,z_2=z,z_3=1,z_4=\infty\}$ respecting the ordering $z_1<z_2<z_3<z_4$.\footnote{Note that this hinges on whether the $SU(1,1)_q$ acts on $S_q^2$ the same way as its classical analogue. This is a highly non-trivial assumption and recent developments in non-commutative differential geometry \cite{Podles:1987wd} could perhaps help make these arguments more rigorous \cite{majid2005noncommutative, heckenberger2003podles}; the reader is referred to \cite{Jevicki:1998rr,Steinacker:1999clg,Steinacker:2001fj} for applications in physics.} With the above assumptions, the authors of \cite{Bernard:1989jq} are led to the following $q$-deformed amplitude
\begin{align}
    \mathcal{A}^{\text{BL}}_4(s,t) &= \lim_{z_4 \to \infty} \int_{0}^{1} d_q z \  z_4^2 \la \phi_1 \dots \phi_4\ra_q = \int_{0}^{1} d_q z \ (x/q)^{\alpha_1\cdot\alpha_2}G_q(z/q|2;\alpha_2\cdot\alpha_3)\nonumber\\
    & = \frac{\Gamma_{q^2}(-s/2-1)\Gamma_{q^2}(-t/2-1)}{\Gamma_{q^2}(-s/2-t/2-2)}~,
    \label{eq:BL_amplitude}
\end{align}
where one immediately sees that the above amplitude is manifestly crossing symmetric, i.e., $\mathcal{A}^{\text{BL}}_4(s,t) = \mathcal{A}^{\text{BL}}_4(t,s)$. However, the main point of contention lies in the relation satisfied by the Mandelstam variables given by $s+t=-4$. 
This is a departure from the expected constraint $s+t+u=-4$ at the level of four-point scattering involving tachyons. 
Such reduced kinematics appear in various scenarios involving scattering processes in (1+1)-dimensional spacetimes, for example, in the scattering of quarks on QCD flux tubes \cite{Dubovsky:2012sh, EliasMiro:2019kyf}.

Interestingly, on the support of this reduced constraint $s+t=-4$, the amplitude $\mathcal{A}^{\text{BL}}_4(s,t)$ (\ref{eq:BL_amplitude}) vanishes identically
\begin{align}
    \mathcal{A}^{\text{BL}}_4(s,t) &= \frac{\Gamma_{q^2}(-s/2-1)\Gamma_{q^2}(-t/2-1)}{\Gamma_{q^2}(-s/2-t/2-2)} \xrightarrow{s+t=-4} 0~.
    \label{eq:vanishingBLamp}
\end{align}
This is due to the fact that $\Gamma_{q^2}(0) \to \infty$ which easily follows from a well-known identity of the $q$-gamma function
\begin{align}
  \Gamma_q (x+1) = [x]_q \Gamma_q (x)~,  
  \label{q-gammaprop}
\end{align}
where $\Gamma_q(x)$ denotes the $q$-gamma function as above and $[x]_q = \frac{1-q^x}{1-q}$ is the $q$-deformed integer. Using the property (\ref{q-gammaprop}) for $x=0$ along with the fact that $\Gamma_{q}(1)=1$ and $[0]_q=0$ establishes that $\Gamma_{q}(0)$ is divergent. For $q<1$,\footnote{This is required for $q$-deformed amplitudes to satisfy unitarity.} one can redefine $q \to q^2$ such that $\Gamma_{q^2}(0) \to \infty$ and consequently the amplitude $\mathcal{A}^{\text{BL}}_4(s,t)$ (\ref{eq:BL_amplitude}) due to Bernard-Leclair \cite{Bernard:1989jq} vanishes identically on the support of the reduced constraint $s+t=-4$.
As we see shall see below, this reduced constraint originates from other known $q$-deformations of $SU(1,1)_q$ as well -- a fact that we pinpoint to the usage of non-invertible maps at the level of Hopf algebras. One might ask whether an analytic continuation of the amplitude $\mathcal{A}^{\text{BL}}_4(s,t)$ to the entire $(s,t,u)$ kinematic phase space reveals a class of legal $q$-amplitudes and we shall explore this possibility for a different amplitude derived later in section \ref{sec:asympproptwistedamp}.

\subsection{Matsuzaki-Suzuki deformation}\label{sec:MS_deformation}
In this subsection, we continue our analysis with non-invertible Hopf algebras and we are led to the analysis carried out by Matsuzaki and Suzuki \cite{Matsuzaki:1991vu}. In comparison to the DJ deformation reviewed above, \cite{Matsuzaki:1991vu} uses a different representation of the $q$-deformed universal enveloping algebra $\mathcal{U}_q(SU(1,1))$ defined by the following Lie algebra structure:
\begin{align}
\left[\mathcal{L}_{+},\mathcal{L}_{-}\right]_q = \frac{1-\mathcal{H}^2}{1-q}~, \qquad q^{\pm 1} \mathcal{H} \mathcal{L}_{\pm} = \mathcal{L}_{\pm} \mathcal{H}~,
\end{align}
where the $q$-commutator bracket is defined as $[A,B]_q \equiv AB - qBA$ and $\mathcal{H}=q^{\mathcal{L}_0}$, such that it reduces to the usual commutation relations in the $q \to 1$ limit. As before, it comes equipped with a Hopf algebra structure $\mathcal{U}_q \rightarrow \mathcal{U}_q \otimes \mathcal{U}_q$ defined by the $q$-comultiplication map $\Delta_q$:
\begin{align}\label{eq:hopf_algebra_of_MS}
\Delta_q(\mathcal{L}_{\pm}) = \mathcal{L}_{\pm} \otimes \mathbb{I} + \mathcal{H} \otimes \mathcal{L}_{\pm}~, \quad \Delta_q(\mathcal{H}) = \mathcal{H} \otimes \mathcal{H}~. 
\end{align}
There again exists a similar representation of $\mathcal{U}_q(SU(1,1))$ as operators on the space of functions $f(z)$:
\begin{align}
\mathcal{L}_{-} f(z) = \frac{1}{z} \frac{f(zq)-f(z)}{1-q}~, \qquad \mathcal{L}_{+} f(z) = z \frac{q^{2\Delta}f(zq)-f(z)}{1-q}~, \qquad \mathcal{H}f(z) = q^{\Delta} f(zq)~.  
\end{align}
We begin with the ansatz considered in \cite{Matsuzaki:1991vu} for the $N$-point correlator
\begin{align}
    \la \phi_1(z_1) \dots \phi_N(z_N)\ra = \prod_{i<j}^{N} z_{i}^{-\gamma_{ij}} {}_{1}\varphi_{0} \left(\gamma_{ij};~x_{ij} q^{a_{ij}}\right)~,
    \label{N-ptansatz}
\end{align}
where $a_{ij}, \gamma_{ij}$ are constants that are to be determined by the application of $q$-Ward identities. The $q$-hypergeometric (or the basic hypergeometric) function ${}_{1}\varphi_{0}$ in (\ref{N-ptansatz}) is defined as\footnote{These are related to the two-point functions for the DJ deformation in the previous section by
\begin{equation*}
    {}_{1}\varphi_{0}(\gamma;~q^{a}x) = G_{\sqrt{q}}\left(x|~a; \gamma\right)~,
\end{equation*}
~~~~~and consequently the observables in the DJ and MS deformations are related by a $q^2\to q$ redefinition.}
\begin{align}
{}_{1}\varphi_{0}(\gamma;~x) = \prod_{l=0}^{\infty} \frac{1-xq^{\gamma+l}}{1-xq^l}~, 
\label{eq:MS_ansatz_seed}
\end{align}
where it is worth noting that the seed of the ansatz in the DJ deformation (\ref{eq:DJ_ansatz_seed}) differs from the basic hypergeometric seed above. This basic hypergeometric function ${}_{1}\varphi_{0}$ satisfies the $q$-difference equation given by:
\begin{align}
{}_{1}\varphi_{0}(\gamma;~x q^{a-1}) = \frac{1-x q^{a+\gamma-1}}{1-x q^{a-1}} {}_{1}\varphi_{0}(\gamma;~x q^{a})~.   
\end{align}
It is worth stressing that while the two- and three- point solutions were originally derived in \cite{Matsuzaki:1991vu}, the four-point solution is still lacking and is derived in the discussion below. 

The application of the $q$-Ward identity due to $\mathcal{H}$ from the $(N-1)$-fold coproduct leads to the same constraint as (\ref{eq:H0_Ward_identity}) upto an overall minus sign. The general $q$-Ward identities for the other two generators in the above representation when applied to the $N$-point ansatz is given by
\begin{align}
    \Delta^{(N-1)}_q(\mathcal{L}_{n}) &= \sum_{i=1}^{N}q^{\sum_{j=1}^{i-1}\Delta_j}\la\phi_1(qz_1)\dots \phi_{i-1}(qz_{i-1})\hat{\phi}_{i}(z_i)\phi_{i+1}(z_{i+1})\dots\phi_N(z_N)\ra~,
\label{ActionofL_N}
\end{align}
where we have defined $\hat{\phi}(z) = z^{n+1}[\partial]_q\phi(z)+z^n[(n+1)h]_q\phi(qz)$. Now, the action of the $\mathcal{L}_{-}$ generator applied to the $N=4$ case using (\ref{N-ptansatz}) yields the following constraints on the undetermined parameters: 
\begin{align}
a_{12} &= 1-\Delta_1~,~a_{23} = 1-\Delta_2~,~a_{34} = 1-\Delta_3~,~a_{14}+\gamma_{14} = 1+\Delta_4~,~\gamma_{13} = 0 =\gamma_{24}~. \label{eq:Lm_constraints_MS}
\end{align}
Analogously, the above $q$-Ward identity for the $\mathcal{L}_{+}$ generator leads to the following additional constraints:
\begin{align}
\gamma_{14}+\gamma_{34} = \Delta_3+\Delta_4~,~ \gamma_{12}+\gamma_{23} = 2\Delta~,~\gamma_{34}+\gamma_{23} = \Delta_3+\Delta_4~,~\gamma_{12}+\gamma_{14} = 2 \Delta~, \label{Lpconstraints}
\end{align}
where we have utilized the additional Ward identity constraint $\Delta_1=\Delta_2 = \Delta$ to simplify the above four-point solution. Employing the same procedure as in section \ref{sec: DJ_deformation}, one can obtain the amplitude for this particular deformation to be
\begin{equation}
    \mathcal{A}^{\text{MS}}_4(s,t)
    = \frac{\Gamma_{q}(-s/2-1)\Gamma_{q}(-t/2-1)}{\Gamma_{q}(-s/2-t/2-2)}~.
    \label{eq:MS_amplitude}
\end{equation}
One immediately notices that this deformed amplitude also vanishes identically on the support of the reduced constraint $s+t=-4$, based on arguments presented below (\ref{eq:vanishingBLamp}). 
\subsection{Woronowicz deformation}
\label{sec:woronowiczdeformation}
We now move on to another example of a non-invertible $q$-deformation of the conformal group, namely the Woronowicz deformation \cite{Woronowicz:1987wr} and follow a similar construction for the corresponding $q$-CFT. In such a deformation, the $SU(1,1)_q$ quantum Lie algebra is defined as 
\begin{align}
    [V_0,V_{\pm}]_{q^2} &\equiv q^2V_0V_{\pm}-\frac{1}{q^2}V_{\pm}V_0= \pm V_{\pm}~, \nonumber \\
    [V_+,V_{-}]_{q^{-1}} &\equiv q^{-1}V_+V_- -qV_{-}V_{+} = V_0~,
\end{align}
where we have introduced the general $q$-Lie bracket considered in \cite{Bernard:1989jq,Zachos:1990wi, Kulish1983QuantumLP} and is defined as 
    $[A,B]_{(q,p)} \equiv qA B-pB A$,
with the further identification $[\cdot,\cdot]_{(q^n,q^{-n})} \equiv [\cdot,\cdot]_{q^n}$. The comultiplication map in this deformation for the corresponding generators of the group algebra is defined to be
\begin{align}
\Delta_W(V_{0}) &= V_{0}\otimes \mathds{1}+\bigg[\mathds{1}-\bigg(q^2-\frac{1}{q^2}V_0\bigg)\bigg]^{1/2}\otimes V_{0}~, \nonumber \\
\Delta_W(V_{\pm}) &= V_{\pm}\otimes \mathds{1}+\bigg[\mathds{1}-\bigg(q^2-\frac{1}{q^2}V_0\bigg)\bigg]\otimes V_{\pm}\label{eq:Hopf_algebra_Woronowicz}~.
\end{align}
These generators are related to the generators of the DJ algebra defined in (\ref{eq:DJ_algebra}) via the following invertible map
\begin{align}
    V_{\pm} = \sqrt{\frac{2q}{q+q^{-1}}}\mathcal{H}^{-1}\mathcal{L}_{\pm}~,~V_0 = \frac{1-\mathcal{H}^{-4}}{q^2-q^{-2}}~\label{eq:V0_transformation}~,
\end{align}
as a result of which the Hopf algebra (\ref{eq:Hopf_algebra_Woronowicz}) can be rewritten in terms of the generators of the DJ algebra as 
\begin{align}
    \Delta_W(V_{0}) &= \frac{1-\mathcal{H}^{-4}}{q^2-q^{-2}}\otimes \mathds{1}+\mathcal{H}^{-4}\otimes \frac{1-\mathcal{H}^{-4}}{q^2-q^{-2}}~,\\
    \Delta_W(V_{\pm}) &= \sqrt{\frac{2q}{q+q^{-1}}}\mathcal{H}^{-1}\mathcal{L}_{\pm}\otimes \mathds{1}+\mathcal{H}^{-2}\otimes \sqrt{\frac{2q}{q+q^{-1}}}\mathcal{H}^{-1}\mathcal{L}_{\pm}~.
\end{align}
Taking inspiration from the ansatz considered in sections \ref{sec: DJ_deformation} and \ref{sec:MS_deformation} for the DJ and MS deformations, respectively, we write the $N$-point ansatz for the Woronowicz deformation as
\begin{equation}
    \la\phi_{\Delta_1}(z_1)\dots\phi_{\Delta_N}(z_N)\ra = \prod_{i<j}z_i^{\gamma_{ij}}f_q(\{x_{ij}\}|\{a_{ij};\gamma_{ij}\})\label{eq:Woronowicz_Npt_Ansatz}~.
\end{equation}
As before, the $N$-point $q$-Ward identity for $V_0$ is given by the $(N-1)$-fold coproduct
\begin{align}
    \Delta_{W}^{(N-1)}(V_0)\la\phi_{\Delta_1}(z_1)\dots\phi_{\Delta_N}(z_N)\ra &= q^{4\left(\sum_{i<j}\gamma_{ij}-\sum_{i=1}^N\Delta_i\right)}\la\phi_{\Delta_1}(q^{-4}z_1)\dots\phi_{\Delta_N}(q^{-4}z_N)\ra \nonumber \\
    &= \la\phi_{\Delta_1}(z_1)\dots\phi_{\Delta_N}(z_N)\ra~,
\end{align}
which then imposes a constraint on the $\gamma_{ij}$'s in terms of the $\Delta_i$'s, identical to the previously derived relation (\ref{eq:H0_Ward_identity}) for the DJ and MS deformations. The two-point $q$-Ward identity then takes the following form for $V_{-}$
\begin{align}
    &z_1q^{-\Delta_1}\left(q^{2\Delta_1}\la\phi_{\Delta_1}(z_1)\phi_{\Delta_2}(z_2)\ra-q^{-2\Delta_1}\la\phi_{\Delta_1}(q^{-2}z_1)\phi_{\Delta_2}(z_2)\ra\right)\nonumber\\
    &-z_2q^{-2\Delta_1-\Delta_2}\left(q^{2\Delta_2}\la\phi_{\Delta_1}(q^{-2}z_1)\phi_{\Delta_2}(q^{-2}z_2)\ra-q^{-2\Delta_2}\la\phi_{q^{-2}\Delta_1}(q^{-2}z_1)\phi_{\Delta_2}(z_2)\ra\right)=0~,
    \label{eq:2pt_V+_ward_idenity}
\end{align}
while for $V_{+}$ one gets 
\begin{align}
    &\frac{q^{-\Delta_1}}{z_1}\left(\la\phi_{\Delta_1}(z_1)\phi_{\Delta_2}(z_2)\ra-\la\phi_{\Delta_1}(q^{-2}z_1)\phi_{\Delta_2}(z_2)\ra\right)\nonumber\\
    &~~~-\frac{q^{-2\Delta_1-\Delta_2}}{z_2}\left(\la\phi_{\Delta_1}(q^{-2}z_1)\phi_{\Delta_2}(q^{-2}z_2)\ra-\la\phi_{\Delta_1}(q^{-2}z_1)\phi_{\Delta_2}(z_2)\ra\right)=0~.
    \label{eq:2pt_V-_ward_idenity}
\end{align}

For the above $q$-Ward identities to be compatible with one another we have to further impose the constraint $\Delta_1 = \Delta_2$. Then substituting our general ansatz (\ref{eq:Woronowicz_Npt_Ansatz}) into (\ref{eq:2pt_V-_ward_idenity}), one finds the following $q$-difference equation for the unknown $q$-function:
\begin{equation}
    (1-xq^{-2\Delta})f_q(x|a;\gamma) = (1-xq^{2\Delta})f_q(q^2 x|a;\gamma)~.\label{eq:2pt_f_q_DE}
\end{equation}
This is exactly identical to the relation (\ref{eq:Gq_identity}), the unique solution of which is given by $f_q = G_q$ along with $a=2\Delta$. This ensures that the two-point solution for the Woronowicz deformation matches exactly the one for the DJ deformation considered in section \ref{sec: DJ_deformation}. This trend continues even at the level of the three-point function, the solution of which is given by the expression
\begin{align}
    \la\phi_{\Delta_1}(z_1)\phi_{\Delta_2}(z_2)\phi_{\Delta_3}(z_3)\ra =& \prod_{i<j}z_i^{\gamma_{ij}}G_q\left(z_{12}|\gamma_{12};\Delta_1+\Delta_2\right) \nonumber \\
    &\qquad \times G_q\left(z_{13}|\gamma_{13};-2\gamma_{13}-\Delta_1-\Delta_3\right)G_q\left(z_{23}|\gamma_{23};\Delta_2+\Delta_3\right)~,
\end{align}
and $\gamma_{ij} = \Delta_k-\Delta_i-\Delta_j$ for $i\neq j\neq k \in \{1,2,3\}$, which is the same result as the DJ type deformation \cite{Bernard:1989jq}. Unsurprisingly, this is the case even at the level of the four-point correlator in which case our solution exactly matches the relation (\ref{eq:Lm_constraints_BC1}). As we elaborate further in the next section, this may seem obvious from the fact that the Woronowicz deformation is related to the DJ algebra via an \textit{invertible} functional deformation map. This happens to be a group isomorphism as long as $q\neq e^{2\pi i/m}$ for any $m \in \mathbb{N}$, which is when the denominator of (\ref{eq:V0_transformation}) blows up and renders the map ill-defined. These maps have been explicitly constructed in \cite{Zachos:1990wi}.

\subsection{Space of QUE-algebras}
\label{sec:spaceofQUE-algebras}
As of now, we have considered the three most well-known deformations of the $SU(1,1)$ group, namely the Drinfel'd-Jimbo (DJ), Matsuzaki-Suzuki (MS) and the Woronowicz deformation. Observing the similarities between the $n$-point correlators arising in $q$-CFTs constructed out of such deformations, it is natural to ask if there exist certain functional isomorphisms between the corresponding Hopf algebras. Such a functional deformation map $Q$ is indeed well-known and was first constructed by Curtright and Zachos \cite{Curtright:1989sw}. In particular, the $q$-Hopf algebra arising from the DJ deformation of the $SU(1,1)$ Hopf algebra given by (\ref{eq:hopf_algebra_of_DJ}) acts as a \textit{base} algebra\footnote{In what follows, we shall refer to the Drinfel'd-Jimbo (DJ) deformation also as the \textit{base} algebra and use these terminologies interchangeably.} for all other Hopf algebras considered so far, namely (\ref{eq:hopf_algebra_of_MS}) and (\ref{eq:Hopf_algebra_Woronowicz}) \cite{Zachos:1990wi}. More explicitly, these are related as
\begin{equation}
    Q(\Delta(\mathfrak{g})) = U_q^{-1}\Delta_q(\mathfrak{g}_q) \, U_q~,\label{eq:similarity_transform-2}
\end{equation}
where $U_q \in \text{End}(\mathfrak{g}_q\otimes \mathfrak{g}_q,\mathbb{C})$ is a similarity transformation between the two Hopf algebras (for more details, please refer to appendix \ref{app:quantum_group}).
To illustrate an explicit example for the construction of these similarity transforms, we consider the following $SU(1,1)$ group generator in the 2d spin-$\frac{1}{2}$ representation
\begin{equation*}
    j_+ = \frac{1}{2}\begin{pmatrix}
        0 & 1 \\
        0 & 0 
    \end{pmatrix}~.
\end{equation*}
This maps identically to the $SU(1,1)_q$ generators for the base algebra i.e., $\mathcal{L}_+=j_+$. The corresponding coproduct (\ref{eq:hopf_algebra_of_DJ}) in this representation is written as 
\begin{equation}
    \Delta_q(j_+) = \frac{1}{\sqrt{2}}\begin{pmatrix}
        0 & \frac{1}{\sqrt{q}}& \sqrt{q} & 0 \\
        0 & 0 & 0 & \frac{1}{\sqrt{q}} \\
         0 & 0 & 0 &\sqrt{q} \\
         0 & 0 & 0 & 0 
    \end{pmatrix}~.
\end{equation}
Likewise, the corresponding generator in the Woronowicz deformation in the spin $1/2$ representation is 
\begin{equation}
    v_+ = \frac{1}{\sqrt{q+q^{-1}}}\begin{pmatrix}
        0 & 1 \\
        0 & 0
    \end{pmatrix}~,
\end{equation}
and the coproduct given by (\ref{eq:Hopf_algebra_Woronowicz}) in the matrix representation is
\begin{equation}
    Q^{\text{Woronowicz}}(\Delta(j_+)) = \Delta_{W}(v_+) = \frac{1}{\sqrt{1+q^2}}\begin{pmatrix}
        0 & 1 & \sqrt{q} & 0 \\
        0 & 0 & 0 & \sqrt{q}\\
         0 & 0 & 0 & \sqrt{q^3} \\
         0 & 0 & 0 & 0 
\end{pmatrix}~.
\end{equation}
There is a whole family of similarity transformations parameterised by $(z_1,z_2)$ between the two coproducts and are given as follows
\begin{align}
  \label{eq: simmatU} U_q(z_1,z_2) &=
    \begin{pmatrix}
         2 q z_1 & 0 & \sqrt{2}z_2& 0 \\
         0 & q\sqrt{2(1+q^2)}z_1 & 0 & q\sqrt{1+q^2}z_2 \\
         0 & 0 & q\sqrt{2(1+q^2)}z_1 & 0 \\
         0 & 0 & 0 & (1+q^2)z_1 
    \end{pmatrix}~, 
\end{align}
for $q\in \mathbb{R}_{\geq 0}$, and $z_1,z_2 \in \mathbb{C}$. Similar arguments can be made to construct similarity transforms between the other two generators $\{j_0,j_{-}\}$ and 
$\{v_0,v_-\}$. However, as pointed out in \cite{Zachos:1990wi,Curtright:1989sw}, this isomorphism breaks down when $q = \text{root of unity}$, particularly highlighting the lack of invertibility with respect to the classical Hopf algebra and the Lie group. These examples of $q$-deformed Lie groups and Hopf algebras that are related by a similarity transform (\ref{eq:similarity_transform-2}) to the base algebra are not restricted to the ones considered in this section --- there are further known deformations like those introduced by Witten \cite{Witten:1989rw} and Curtright-Zachos \cite{Curtright:1989sw}. These deformations also fall into the category of non-invertible Hopf algebras.\footnote{In the mathematics literature, these are often referred to as ``equivalent" Hopf algebras (see appendix \ref{app:quantum_group} for further details).} A summary of the various Quantum Universal Enveloping (QUE) algebras of $SU(1,1)_q$ discussed in the paper are listed in table \ref{table:generators} along with their associated generators.
\begin{table}
 \centering
\begin{tabularx}{0.8\textwidth} { 
  | >{\raggedright\arraybackslash}X 
  | >{\centering\arraybackslash}X 
  | >{\raggedleft\arraybackslash}X | }
 
 \hline
   Deformation & Generators & Representation \\
  \hline
   Classical ($q=1$) & $\{J_0,J_+,J_-\}$ & arbitrary $j$\\
   & $\{j_0,j_+,j_-\}$ & $j=1/2$ \\ 
             & $\{L_0,L_+,L_-\}$ & $j=1$ \\ 
  \hline
   DJ (base algebra) & $\{\mathcal{L}_0,\mathcal{L}_+,\mathcal{L}_-\}$ & arbitrary $j$\\
  \hline
  Woronowicz & $\{V_0,V_+,V_-\}$ & arbitrary $j$\\ 
             &$\{v_0,v_+,v_-\}$ & $j=1/2$\\  
  \hline
  $\text{Witten}_1$ & $\{E_0,E_+,E_-\}$ & arbitrary $j$ \\ 
  \hline
  $\text{Witten}_2$ & $\{W_0,W_+,W_-\}$ & arbitrary $j$ \\
  \hline
  CZ & $\{Z_0,Z_+,Z_-\}$ & arbitrary $j$ \\
  \hline
\end{tabularx}
\caption{Different QUE algebras of $SU(1,1)_q$ and the notations used for the associated generators.}
\label{table:generators}
\end{table}

\begin{figure}[!ht]
    \centering
    \includegraphics[scale=0.44]{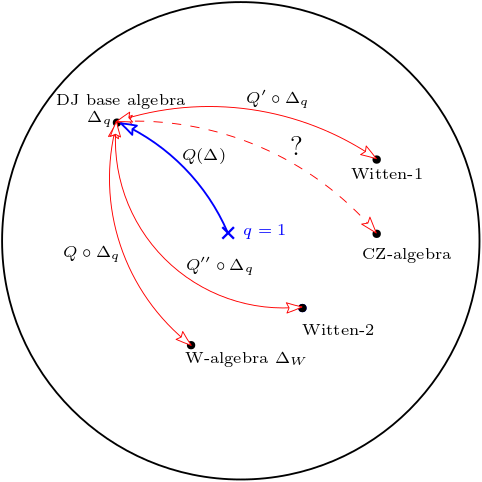}
    \caption{Moduli space of non-invertible Hopf algebras corresponding to QUE groups.}
    \label{fig:space_of_que_algebras}
\end{figure}

The non-invertible nature of the Hopf algebra map when flowing from the classical group $SU(1,1)$ to its quantum counterpart (the \textit{base} algebra deformation studied in section \ref{sec: DJ_deformation}) accompanied by the existence of an invertible deforming functional relating the base algebra with another quantum deformation (in this case with the Woronowicz algebra as shown above) motivates us to probe the interrelationships between other known quantum deformations of $SU(1,1)_q$. In particular, relying on certain heuristic arguments we attempt to provide strong evidence that amplitudes in $q$-deformed CFTs constructed out of non-invertible Hopf algebras will always vanish identically, \textit{independent} of the representation of the quantum group $SU(1,1)_q$ being studied. We try to back such claims by resorting to the quantitative arguments presented for the DJ and Woronowicz deformations above and by studying the moduli-space of Hopf algebras corresponding to QUE groups depicted in figure \ref{fig:space_of_que_algebras}. 
We highlight the salient features of figure \ref{fig:space_of_que_algebras} as follows and in the process attempt to establish the interrelationships between known deformations of the worldsheet conformal group $SU(1,1)_q$:
\begin{itemize}
    \item \underline{\textbf{The origin $q=1$}}\textbf{:} At the origin of the moduli space lies the well-studied classical group $SU(1,1)$ which describes the global worldsheet conformal group in ordinary string theory. As it is important for what follows, it is worth mentioning that the origin of the moduli space described by $q=1$ trivially sits at a root of unity of $q$ while the other $q$-deformations in the figure do not.
    \item \underline{\textbf{The flow from the origin to the base algebra} $\Delta \xrightarrow{Q(\Delta)} \Delta_q$}\textbf{:}
    We now consider the flow from the origin of the moduli space $q=1$ to the base deformation discussed in section \ref{sec: DJ_deformation}. Although invertible at the level of Lie algebras, the map $Q(\Delta)$ relating their Hopf algebras are non-invertible (and hence depicted by a one-way arrow in figure \ref{fig:space_of_que_algebras}) due to the fact that the classical Hopf algebra of $SU(1,1)$ sits at a root of unity, as emphasized above. The four-point amplitude constructed using the coproduct map $\Delta_q$ associated to the generators of the DJ deformation (\ref{eq:hopf_algebra_of_DJ}) vanishes identically (\ref{eq:vanishingBLamp}) on the support of the reduced constraint $s+t=-4$. On further inspection, we shall argue that this is a feature that persists along flows between different quantum group deformations in the moduli space. 
    \item \underline{\textbf{The flow from the base algebra to the Woronowicz algebra} $\Delta_q \xrightarrow{Q \cdot \Delta_q} \Delta_W$}\textbf{:}\\The map relating the Hopf algebras of the base and the Woronowicz deformation is described by the functional deformation (\ref{eq:similarity_transform-2}), which is a similarity transform explicitly realized by the matrix (\ref{eq: simmatU}). As expounded in \cite{Zachos:1990wi}, such functional deforming maps are invertible and hence denoted by a two-way arrow in figure \ref{fig:space_of_que_algebras}. The invertibility of such maps leads to the suspicion that the set of observables in both the theories would be isomorphic. As explicitly worked out in section \ref{sec:woronowiczdeformation}, the four-point correlators (and consequently the amplitudes) arising out of the Woronowicz deformation is exactly identical to those derived for the base algebra. This concludes that the four-point amplitude of the $q$-CFT based on the Woronowicz deformation vanishes identically. 
    \item \underline{\textbf{The flow from the base algebra to other deformations} $\Delta_q \xrightarrow{\{Q^{\prime}, Q^{\prime \prime}, \dots \} \cdot \Delta_q} \Delta_x$}\textbf{:} \\
    We now turn our attention to other possible flows in the moduli space between the base algebra and other possible deformations of $SU(1,1)_q$ via invertible functional deformations as the one above. We have done an explicit check for the previous flow ($\text{base} \rightarrow \text{Woronowicz}$) that the set of observables in both deformations coincide. However, this coincidence is due to the fact that the functional deforming maps between the base algebra and other deformations are invertible and that they occupy points in the moduli space that are $q \neq \text{root of unity}$. This invertible nature implies that \textit{any} quantum group one flows to from the base algebra will have identical correlation functions while the amplitudes of such theories will vanish. 
    \item \underline{\textbf{The dotted flow from the base algebra to the CZ algebra:}} The algebra proposed by Curtright and Zachos (CZ) in \cite{Curtright:1989sw} is defined by the operators $Z_m$
    \begin{align}
        Z_m = x^{-m} \frac{q^{2 x \partial}-1}{q-q^{-1}}~,
    \end{align}
    where they satisfy the $q$-deformation of the centreless Virasoro algebra 
    \begin{align}
        [Z_m,Z_n]_{q^{n-m}} = [m-n]_q Z_{m+n}~,
    \end{align}
    and the subset of operators $\{Z_1, Z_0, Z_{-1}\}$ comprise the generators of the global conformal group $SU(1,1)_q$. However, despite attempts in constructing a coproduct map for such a deformation \cite{Oh:1992hn}, the Hopf algebra structure is still unknown and hence denoted by a dotted line in figure \ref{fig:space_of_que_algebras}. We note that our analysis, therefore, does not extend to this particular $q$-deformation defined by the CZ algebra. 
\end{itemize}
Using the set of arguments above, both quantitative and inductive, we conclude that the amplitudes constructed out of non-invertible Hopf algebras (in flowing from the classical to the quantum groups) will be vanishing in nature due to the reduced constraint $s+t=-4$. This leads us to pose the following two questions: i) Does there exist a map that is \textit{invertible} at the level of Hopf algebras in flowing from the classical $SU(1,1)$ to the base algebra $SU(1,1)_q$? ii) If yes, what is the set of observables that such invertible Hopf algebras lead to? These questions form the basis of our analysis in the next section.

\section{\texorpdfstring{Towards a $q$-CFT construction from Drinfel'd twists}{Towards a q-CFT construction from Drinfel'd twists}} \label{sec:qCFTDrinfel'd}
Having outlined the construction of observables originating from non-invertible Hopf algebras and their status in the space of QUE algebras, we are now led to the next stage of our analysis. If we conjecture that the origin of the reduced constraint $s+t=-4$ (which is a feature of all $q$-deformations considered so far) stems from the non-invertibility of Hopf algebras, we are naturally led to the following question -- do functional deformations exist that have well-defined, invertible coproduct maps? The answer is \textit{yes} and is provided by the mathematical objects called Drinfel'd twists \cite{drinfel'd1989quasi}. These Drinfel'd twists provide us with an invertible map that holds at the level of (quasi-triangular) Hopf-algebras as one flows from undeformed $SU(1,1)$ to the base alegbra $SU(1,1)_q$. We explore the definition of such invertible twists (please refer to appendix \ref{app:Drinfel'd_twist} for more formal definitions and mathematical details of the Drinfel'd twist) and use them to construct $q$-CFT correlators via the application of coproduct maps and \textit{twisted} Ward identities.


\subsection{Drinfel'd twists and twisted Ward identities}
The Drinfel'd twisted coproduct is defined as \cite{drinfeld1990quasi,1997grgp.conf..293D}
\begin{equation}
    \tilde{\Delta}_h(x) = \mathcal{F}\Delta(x)\mathcal{F}^{-1}~,
\end{equation}
where $\mathcal{F}$ is the twist and is defined as a perturbative expansion in the parameter $h$ such that $q=e^h$. The universal twist element $\mathcal{F}$ is known to have been constructed up to the second order in $h$ \cite{1997grgp.conf..293D}:
\begin{equation}
    \mathcal{F} = \mathcal{F}_0+h\mathcal{F}_1+h^2\mathcal{F}_2+\mathcal{O}(h^3)~.
\end{equation}
In the subsequent analysis, we will only work with the leading order correction in $h$ to the twist $\mathcal{F}$ defined above, following which the $(N-1)$-fold coproduct map can be expanded as
\begin{equation}
    \tilde{\Delta}^{(N-1)}_h = \Delta^{(N-1)}_0+h\Delta^{(N-1)}_1+\mathcal{O}(h^2)~.
    \label{eq:N-foldtwistedcoproduct}
\end{equation}
Henceforth, we shall set $\Delta^{(N-1)}_0 \equiv \Delta^{(N-1)}$. Likewise, the correlators can be expanded as follows
\begin{equation}
    \la\phi_{1}\dots\phi_N\ra_h = \la\phi_{1}\dots\phi_N\ra^{(0)}+h\la\phi_{1}\dots\phi_N\ra^{(1)}+\mathcal{O}(h^2)~,
\end{equation}
where we have defined $\phi_i \equiv \phi_{\Delta_i}(z_i)$. Applying this $(N-1)$-fold coproduct map (\ref{eq:N-foldtwistedcoproduct}) to the $N$-point correlators gives rise to what we call the $N$-point ``\textit{twisted}'' Ward identities associated to the Drinfel'd twists introduced above:
\begin{align}
    \tilde{\Delta}^{(N-1)}_h(L_{n}) \la\phi_{1}\dots\phi_N\ra_h &= \Delta^{(N-1)}(L_{n})\la\phi_{1}\dots\phi_N\ra^{(0)} \nonumber \\
    &+h\left(\Delta^{(N-1)}(L_{n})\la\phi_{1}\dots\phi_N\ra^{(1)}+\Delta^{(N-1)}_1(L_{n})\la\phi_{1}\dots\phi_N\ra^{(0)}\right) + \mathcal{O}(h^2) =0~,
    \label{twistedWardiden}
\end{align}
for $n\in\{-1,0,1\}$. Using the fact that
\begin{align}
\Delta^{(N-1)}(L_{n})\la\phi_{1}\dots\phi_N\ra^{(0)} = \sum_{i=1}^N\la\phi_{1}\dots L_n\phi_i\dots\phi_N\ra^{(0)}~,
\label{eq:N-foldonzerothorder}
\end{align}
along with the following action of the Hopf algebra at $\mathcal{O}(h)$ (please refer to appendix \ref{app:Drinfel'd_twist} for a short discussion on the $\mathcal{O}(h)$ twisted coproduct, while a detailed derivation of the same can be found in \cite{1997grgp.conf..293D}): 
\begin{equation}
    \Delta_1(L_{n}) \equiv [\mathcal{F}_1,\Delta(L_n)] = L_n\otimes L_0-L_0\otimes L_n~,
    \label{eq:orderonetwistedcoproduct}
\end{equation}
leads us to the following two-point ($N=2$) twisted Ward identity
\begin{align}
    \la L_{n}(\phi_1)\phi_2\ra^{(0)}+\la\phi_1 L_{n}(\phi_2)\ra^{(0)} +&h\left(\la L_{n}(\phi_1)L_0(\phi_2)\ra^{(0)}-\la L_0(\phi_1)L_{n}(\phi_2)\ra^{(0)}\right.\nonumber\\
    &\qquad \qquad \qquad \qquad +\left.\la L_{n}(\phi_1)\phi_2\ra^{(1)}+\la\phi_1 L_{n}(\phi_2)\ra^{(1)}\right) + \mathcal{O}(h^2)= 0~.
    \label{two-point_twisted_ward_identity}
\end{align}
Further, using the $(N-1)$-fold coproduct map (\ref{eq:N-foldcoproduct}) adapted to $N=3$, along with relations (\ref{eq:N-foldonzerothorder}) and (\ref{eq:orderonetwistedcoproduct}) leads us to the three-point twisted Ward identity
\begin{align}
     \sum_{i=1}^3\la \phi_1\dots L_{n}\phi_i\dots\phi_3&\ra^{(0)} +h\bigg(\la L_{n}(\phi_1)L_0(\phi_2)\phi_3\ra^{(0)}\nonumber-\la L_0(\phi_1)L_{n}(\phi_2)\phi_3\ra^{(0)}\\
     &\qquad \quad+\la L_n(\phi_1)\phi_2L_{0}(\phi_3)\ra^{(0)}-\la \phi_1 L_0(\phi_2)L_{n}(\phi_3)\ra^{(0)}+\la \phi_1 L_n(\phi_2)L_{0}(\phi_3)\ra^{(0)}\nonumber\\
     &\qquad \qquad  -\la L_0(\phi_1) \phi_2L_{n}(\phi_3)\ra^{(0)}+\sum_{i=1}^3\la \phi_1\dots L_{n}\phi_i\dots\phi_3\ra^{(1)}\bigg)+\mathcal{O}(h^2)= 0~.
\end{align}
Generalizing the above relations, the twisted Ward identity at $N$-point can be expressed compactly as follows
\begin{align}
    \sum_{i=1}^{N}\la \phi_1\dots L_{n}(\phi_i)\dots\phi_N\ra^{(0)}+h&\left(\sum_{m\neq k}\text{sgn}(m-k)\la \phi_1\dots L_{n}(\phi_k)\dots L_0(\phi_m)\dots\phi_N\ra^{(0)}\right.\nonumber\\
    &\qquad \qquad \qquad  \left.+\sum_{i=1}^{N}\la \phi_1\dots L_{n}(\phi_i)\dots\phi_N\ra^{(1)}\right)+\mathcal{O}(h^2)=0~.
\end{align}
\subsection{Two-point twisted correlators}

We begin with a two-point ansatz that matches with the conventional CFT two-point correlator at the leading order, along with a $\mathcal{O}(h)$ correction given by
\begin{align}
   \la \phi_{\Delta_1}(z_1) \phi_{\Delta_2}(z_2) \ra_h = \frac{\mathcal{N}_{12}}{z_{12}^{2\Delta_1}} + h f(z_1,z_2; \Delta_1,\Delta_2) + \mathcal{O}(h^2)~,  
\end{align}
where $f(z_1,z_2; \Delta_1,\Delta_2)$ is a correction to be determined by the application of twisted Ward identities. We shall be working with the differential representation for the undeformed $SU(1,1)$ generators $L_{n}^{(i)}$ given by
\begin{align}
    L_0^{(i)} = z_i \partial_{z_i} + \Delta_i~, \quad L_{+}^{(i)} = z_i^2 \partial_{z_i} + 2 \Delta_i z_i~, \quad  L_{-}^{(i)} = \partial_{z_i}~.
\end{align}
We now turn our attention to solving for the unknown function $f(z_1,z_2; \Delta_1,\Delta_2)$ by application of the two-point twisted Ward identity (\ref{two-point_twisted_ward_identity}) corresponding to $L_{-}$. While this is trivially satisfied at the leading order, one obtains the following differential equation for $f(z_1,z_2; \Delta_1,\Delta_2)$ at the sub-leading order in $h$
    \begin{align}
        (\partial_{z_1} + \partial_{z_2})f(z_1,z_2; \Delta_1,\Delta_2) = -\frac{2\Delta_1}{z_{12}^{2\Delta_1+1}}~.
    \end{align}
Solving the above PDE leads to the solution
\begin{align}
    f (z_1,z_2; \Delta_1,\Delta_2) = -\frac{2\Delta_1 z_1}{z_{12}^{2\Delta_1+1}} + \frac{C_{12}(\Delta_1)}{z_{12}^{2\Delta_1}}~,
\end{align}    
where $C_{ij}(\{\Delta_i\})$ is simply a constant of integration which contributes to the overall normalisation of the two-point function and is dropped from our analysis from here on.\footnote{It can also be checked that it drops out even at the level of the three-point function and higher.} Setting the overall normalization $\mathcal{N}_{12}$ to unity leads us to the following twisted two-point function
\begin{align}
  \la \phi_i \phi_j \ra_h = \delta_{\Delta_i,\Delta_j}\left(\frac{1}{z_{ij}^{2\Delta_i}} + h \, \frac{(-2\Delta_i) z_i}{z_{ij}^{2\Delta_i+1}}\right)~. 
  \label{2-ptfunc}
\end{align}
It is worth noting that the action of the twisted Ward identities due to $L_{+}, L_0$ are trivially satisfied by the two-point solution (\ref{2-ptfunc}) at both orders $\mathcal{O}(1), \mathcal{O}(h)$. This is made possible by the fact that the conformal dimensions are equal $\Delta_1=\Delta_2$, which essentially arises from the leading order contribution of the $L_{+}$ Ward identity. 

\subsection{Three-point twisted correlators}
At the level of higher-point correlators, we generalize our two-point ansatz to write it in the following form
\begin{align}
    \la \phi_{\Delta_1}(z_1) \dots \phi_{\Delta_n}(z_n) \ra_h = \prod_{i<j} z_{ij}^{-\gamma_{ij}} \left(1+h \sum_{i<j} \frac{a_{ij}}{z_{ij}} z_i\right)~,
    \label{npttwistedansatz}
\end{align}
where the constants $a_{ij}(\{\gamma_{ij},\Delta_i\})$ are to be determined by the action of the twisted Ward identities. At leading order, the twisted Ward identity corresponding to $L_{-}$ leads us to the following constraints on the $\gamma$'s in terms of the conformal dimensions $\Delta_i$
\begin{align}
    \gamma_{ij} = \Delta_i + \Delta_j - \Delta_k~, \qquad \forall \, i \neq j \neq k~, 
    \label{3ptorderh0sol}
\end{align} 
while the corresponding $\mathcal{O}(h)$ twisted Ward identity takes the form
     \begin{equation} \sum_{m=1}^3\partial_{m}\la\phi_1\dots\phi_m\dots\phi_3\ra^{(1)}+\sum_{m\neq k}\text{sgn}(m-k)\left(z_m\partial_m\partial_k+h_m\partial_k\right)\la\phi_1\dots\phi_k\dots\phi_m\dots\phi_3\ra^{(0)}=0
     \end{equation}
     where $\partial_{i} \equiv \partial_{z_i}$. Making use of our ansatz above (\ref{npttwistedansatz}) for the case of the three-point correlator, we get 
     \begin{equation}
                  \prod_{i<j}z_{ij}^{-\gamma_{ij}}\left(\frac{1}{z_{12}}\left(\gamma_{12}\left(\Delta_3-1\right)+a_{12}\right)+\frac{1}{z_{23}}\left(\gamma_{23}\left(\Delta_1-1\right)+a_{23}\right)-\frac{1}{z_{13}}\left(\gamma_{13}\left(\Delta_2+1\right)-a_{13}\right)\right)=0~,
     \end{equation}
     which leads to the following constraints on the unknowns
     \begin{equation}
         a_{ij} = \gamma_{ij}(1+(-1)^{i+j}\Delta_k) ~,\quad \forall~k\neq i\neq j~.\label{3ptorderhsol}
     \end{equation}
The remaining twisted Ward identity corresponding to $L_{0}$ leads us to the expected constraint as follows:
\begin{align}
    \sum_{i<j} \gamma_{ij} = \sum_{i=1}^4 \Delta_i~,
    \label{L0_constraint}
\end{align}
while the corresponding action due to $L_{+}$ is trivially satisfied by the three-point solution at both orders $\mathcal{O}(1), \mathcal{O}(h)$. Combining both the leading (\ref{3ptorderh0sol}) and sub-leading (\ref{3ptorderhsol}) terms in $h$, we arrive at the following final solution for the three-point twisted correlator in terms of the conformal dimensions:
\begin{align}
a_{12} = (\Delta_1+\Delta_2-\Delta_3)(1-\Delta_3)~, ~~ a_{13} = (\Delta_1+\Delta_3-\Delta_2)(\Delta_2+1)~, ~~ a_{23} = (\Delta_2+\Delta_3-\Delta_1)(1-\Delta_1)~. \label{3-pttwistedsolution}   
\end{align}

\subsection{Four-point twisted correlators and amplitudes}
\label{sec:twsitedamp}
\subsubsection{Twisted four-point correlator}
We work with our $N$-point ansatz (\ref{npttwistedansatz}) adapted to $N=4$ and determine the undetermined constants as before by the application of twisted Ward identities. At the leading order, the twisted Ward identity corresponding to $L_{+}$ leads us to the following constraints on $\gamma$'s in terms of the conformal dimensions $\Delta_i$:
\begin{align}
    \sum_{i \neq j} \gamma_{ij} = 2 \Delta_i ~,
    \label{4ptorderh0sol}
\end{align} 
where we note that $\gamma_{ij}=\gamma_{ji}$ for the above formula to hold. 
At the sub-leading order in $h$, the four-point twisted Ward identity leads us to the following solution for the four-point correlator:
   \begin{align}
a_{12} &= -\gamma_{12} (\gamma_{12}+1 -2\Delta)~,~a_{14} = \gamma_{14} (\gamma_{14}-1-2\Delta)~,~a_{23} = -\gamma_{23} (\gamma_{23}+1-2\Delta)~, \nonumber \\
    a_{34} &= -\gamma_{34} (\gamma_{34}+1-2\Delta)~,~a_{24}=0=a_{13}~,
   \end{align}
   which follows from the fact that  $\gamma_{13}=0=\gamma_{24}$ and $\Delta_i = \Delta$. 
It is worth noting that while the action of $L_{0}$ leads us to the expected constraint (\ref{L0_constraint}), the twisted Ward identity corresponding to $L_{-}$ is trivially satisfied. 

\subsubsection{Twisted four-point amplitude}
Given the above solution, one can obtain the corresponding four-point amplitude where we use the parameterisation and the conventional open string gauge-fixing conditions explained in section \ref{sec: DJ_deformation}, along with the fact that we set $\Delta_i=\Delta=1$. Evaluating the integral, one obtains for the first-order correction in $h$ to be 
\begin{align}\label{eq:BD_amplitude}
\mathcal{A}^{\text{twisted}}_4(s,t) &= \lim_{z_1 \to \infty} \int_0^1 d_qx \, z_1^2 \la \phi_{1}(z_1) \phi_{1}(1) \phi_{1}(x) \phi_1(0) \ra_h \nonumber \\
&= B\left(-\frac{s}{2}-1,-\frac{t}{2}-1\right)\left(1+h\frac{st}{4}\right) = \mathcal{A}_4^{V}(s,t)\left(1+h\frac{st}{4}\right)~.
\end{align}
where the Mandelstam variables interestingly again satisfy the reduced constraint $s+t=-4$ and the beta function prefactor $B\left(-s/2-1,-t/2-1\right)$ is identified as the well-known Veneziano amplitude $\mathcal{A}_4^{V}(s,t)$.
Despite being crossing-symmetric in $s,t$, the twisted amplitude $\mathcal{A}^{\text{twisted}}_4(s,t)$ vanishes identically on the support of the reduced constraint $s+t=-4$: 
\begin{align}
    \mathcal{A}^{\text{twisted}}_4(s,t) = \mathcal{A}^{V}_4(s,t) \left(1+h \frac{st}{4}\right) = \frac{\Gamma(-s/2-1) \Gamma(-t/2-1)}{\Gamma(-s/2-t/2-2)}\left(1+h \frac{st}{4}\right) \xrightarrow{s+t=-4} 0~.
    \label{eq:twistedvanishing}
\end{align}
This leads to a striking result! Even the above amplitude $\mathcal{A}^{\text{twisted}}_4(s,t)$ constructed using invertible Hopf algebra maps (i.e., using the Drinfel'd twists) vanishes identically due to the reduced constraint $s+t=-4$. 
It is worth mentioning that any higher-order corrections to the twisted amplitude $\mathcal{A}^{\text{twisted}}_4(s,t)$ (\ref{eq:BD_amplitude}) coming from the perturbative expansion of the Drinfel'd twist (\ref{eq:N-foldtwistedcoproduct}) will not affect the above conclusion. This is due to the fact that vanishing of the twisted amplitude (\ref{eq:twistedvanishing}) relies on the denominator of the Veneziano amplitude $\mathcal{A}_4^V(s,t)$ being singular. This occurs because the argument of the relevant gamma function $\Gamma(-s/2-t/2-2)$ vanishes on the support of $s+t=-4$ thereby hitting the pole at zero. 

The above finding leads us to conclude that the four-point tachyonic amplitude constructed out of an invertible Hopf algebra (i.e., the Drinfel'd twist) produce twisted amplitudes that vanish due to the reduced constraint $s+t=-4$. Combining this with the results and arguments presented in section \ref{sec:spaceofQUE-algebras} for the space of QUE-algebras (with non-invertible Hopf algebras), we are led to conclude that the tachyonic $S$-matrices for $q$-deformed CFTs with an underlying conformal group $SU(1,1)_q$ are \textit{trivial}, i.e., they are simply identity matrices. This statement also hinges on the fact that the three-point, tachyonic gauge-fixed amplitudes also vanish, which can be seen from the three-point twisted solution (\ref{L0_constraint}) and (\ref{3-pttwistedsolution}) supplemented by the tachyonic condition $\Delta_i = \Delta =1$. Thus, the vanishing of the three- and four-point amplitudes lead us to conclude that all higher-point tachyonic amplitudes vanish in such $q$-deformed worldsheet CFTs rendering their $S$-matrices trivial.  

\section{The dual-resonant twisted amplitude(s)}
\label{sec:asympproptwistedamp}
Due to the vanishing of amplitudes in such theories, it is worth asking whether one should consider moving away from the realm of $q$-CFTs to obtain well-defined, non-vanishing amplitudes. 
Since the source of the vanishing of such $q$-deformed amplitudes lies in the reduced constraint $s+t=-4$, it is interesting to ask whether an analytic continuation into the entire $(s,t,u)$ kinematic phase space defines a valid class of amplitudes. In such a scenario, on grounds of crossing symmetry and polynomial boundedness, the amplitude is expected to have additional terms at $\mathcal{O}(h)$ which we denote by the polynomials $f(s+t)$, $g(s+t)$ and $h(st)$. The $q$-deformed amplitude then takes the general form as follows
\begin{align}
    \mathcal{A}^{\text{twisted,AC}}_4(s,t;\alpha,\beta,\gamma) = \mathcal{A}^{V}_4(s,t) \left[1+h \left(\frac{st}{4} + \alpha f(s+t)st + \beta g(s+t)+\gamma h(st) \right)+\mathcal{O}(h^2)\right]~,
    \label{eq:analcontampmostgen}
\end{align}
with $\alpha$, $\beta$ and $\gamma$ being multiplicative constants. For the above to be a consistent analytic continuation of (\ref{eq:BD_amplitude}), we must recover the latter on the kinematic slicing $u=0$ of the former. For $u=0$, the analytically continued amplitude takes the form
\begin{equation}
\mathcal{A}^{\text{twisted,AC}}_4(s,t;\alpha,\beta,\gamma)\big|_{s+t=-4} = \mathcal{A}^{V}_4(s,t) \left[1+h \left(\frac{st}{4} + \alpha f(-4)st + \beta g(-4)+\gamma h(s) \right)+\mathcal{O}(h^2)\right]~,
    \label{eq:analcontampmostgen-2}
\end{equation}
and one can always redefine the polynomials $f$ and $g$ such that $f(-4)=0=g(-4)$, but for the above to match (\ref{eq:BD_amplitude}), we must have $\gamma=0$.\footnote{In general, one may further add polynomials of the form $f_n(s^n+t^n)$ for $n\in \mathbb{N}/\{1\}$, which would be justified from the perspective of manifest crossing symmetry and polynomial boundedness. However, these drop out from our ansatz (\ref{eq:analcontampgen}) for exactly the same reason as to why the $h(st)$ term drops out --- they do not lead to a consistent analytic continuation of $\mathcal{A}_4^{\text{twisted}}(s,t)$.} Thus, the most general form for the analytically continued twisted amplitude is given by
\begin{equation}
    \mathcal{A}^{\text{twisted,AC}}_4(s,t;\alpha,\beta) = \mathcal{A}^{V}_4(s,t) \left[1+h \left(\frac{st}{4} + \alpha f(s+t)st + \beta g(s+t)\right)+\mathcal{O}(h^2)\right]~.
    \label{eq:analcontampgen}
\end{equation}
Crucially, the Mandelstam invariants of the analytically continued amplitude (\ref{eq:analcontampgen}) now satisfy the desired four-point constraint $s+t+u=-4$. While we leave the analysis of such class of new amplitudes $\mathcal{A}^{\text{twisted,AC}}_4(s,t;\alpha,\beta)$ as future work, we briefly discuss the amplitude in the regime when $\alpha, \beta =0$ such that
\begin{align}
    \mathcal{A}^{\text{twisted,AC}}_4(s,t) = \mathcal{A}^{V}_4(s,t) \left[1+ h \left(\frac{st}{4} \right)+\mathcal{O}(h^2)\right]~,
    \label{eq:analcontamp}
\end{align}
supplemented by the full constraint $s+t+u=-4$. The amplitude above represents the first-order correction in $h$\footnote{In other words around $q \approx 1$ since we have the relation $q=e^h$.} to the Veneziano amplitude $\mathcal{A}_4^V(s,t)$. 
Remarkably, the entire family of twisted amplitudes described by (\ref{eq:analcontampgen}) is manifestly dual resonant. Consequently, the amplitude $\mathcal{A}^{\text{twisted,AC}}_4(s,t)$ described by (\ref{eq:analcontamp}), which belongs to such a family, is also dual resonant and this is made manifest by its Veneziano prefactor. The dual resonant nature is seen from the fact that these twisted amplitudes satisfy:
\begin{itemize}
    \item \textit{Crossing symmetry}: The class of amplitudes (\ref{eq:analcontampgen}) is manifestly crossing symmetric in the Mandelstam invariants $s,t$ due to the dual resonant Veneziano prefactor $\mathcal{A}_4^V(s,t)$ and the functional form of the polynomials $f$ and $g$. 
    \item \textit{Meromorphicity}: Since $f,g$ are polynomials in the Mandelstam variables (and a monomial in the case of (\ref{eq:analcontamp})), the pole structure is entirely borrowed from its Veneziano prefactor $\mathcal{A}_4^V(s,t)$. Therefore, the family of amplitudes so derived has simple poles at positive integer values of $s,t$.
    \item \textit{Polynomial residues}: Considering the residue of the amplitude at $s=n\in \mathbb{N}$, we have
    \begin{align}
        \text{Res}_{s=n}&\left(\mathcal{A}^{\text{twisted,AC}}_4(s,t;\alpha,\beta)\right) =\nonumber\\ &\text{Res}_{s=n}\left(\mathcal{A}^{V}_4(s,t)\right)\left[1+h \left(\frac{nt}{4} + \alpha f(n+t)nt + \beta g(n+t)\right)+\mathcal{O}(h^2)\right]~,
    \end{align}
    since the residue of the Veneziano amplitude at $s=n$ is a polynomial in $t$, so is the case for the term inside the bracket, and it follows that $\text{Res}_{s=n}\left(\mathcal{A}^{\text{twisted,AC}}_4(s,t;\alpha,\beta)\right)$ is a polynomial in $t$. Crossing symmetry implies this to be also true for the residue at $t=n$. Thus, the analytically continued amplitudes (\ref{eq:analcontampgen}) have polynomial residues. 
\end{itemize}
Notice that we can rewrite (\ref{eq:analcontamp}) as:
    \begin{align}
      \mathcal{A}^{\text{twisted,AC}}_4(s,t) &= \mathcal{A}^{V}_4(s,t) \left[1+ h \left(\left(-\frac{s}{2}-1\right)\left(-\frac{t}{2}-1\right)-\frac{s}{2}-\frac{t}{2}-1\right)+\mathcal{O}(h^2)\right]~,\nonumber\\
        &= \frac{\Gamma\left(-s/2-1\right)\Gamma\left(-t/2-1\right)}{\Gamma(-s/2-t/2-2)} \nonumber \\
        &+ h \left(\frac{\Gamma(-s/2)\Gamma(-t/2-1)}{\Gamma(-s/2-t/2-2)} +\frac{\Gamma(-s/2-1)\Gamma(-t/2)}{\Gamma(-s/2-t/2-2)}+\frac{\Gamma(-s/2)\Gamma(-t/2)}{\Gamma(-s/2-t/2-2)}\right)+\mathcal{O}(h^2)~,
        \label{eq:analcontampsat}
    \end{align}
up to a normalisation that only depends on $h$. 
While the second line follows from the definition of the Veneziano amplitude as a beta function, the $\mathcal{O}(h)$ terms in the third line are interestingly the so-called Veneziano satellites, first mentioned in \cite{Gross:1969db}. 
Our analysis above could potentially shed light regarding a worldsheet realization for these dual resonant satellite terms, which have been posited numerous times in the literature. 
More recently, a generalisation of (\ref{eq:analcontampsat}) has been proposed particularly from the perspective of unitarity in \cite{pirsa_PIRSA:23070040}. The authors showed that there exists non-unique unitary deformations of the Veneziano amplitude, in alignment with our findings.
In addition, the above class of amplitudes have an integer spectrum, consistent with string theory. In \cite{Cheung:2023adk}, the authors bootstrap a more general class of the so-called ``hypergeometric" amplitudes by assuming the integer spectrum as an input. It is natural to then speculate whether our amplitude falls into such a class of amplitudes, but as we will see later in the section, their high-energy behaviours are remarkably different. The question of unitarity of such twisted amplitudes and the sub-sub-leading correction in $h$, i.e, at $\mathcal{O}(h^2)$, using the next order Drinfel'd twist will be addressed in upcoming work. 

Having established that the entire family of twisted amplitudes (\ref{eq:analcontampgen}) are dual resonant, we now focus on the specific twisted amplitude $\mathcal{A}^{\text{twisted,AC}}_4(s,t)$ described by (\ref{eq:analcontamp}). In particular, we discuss its asymptotic properties, both in the high-energy and Regge scattering regimes:
\paragraph{High-energy fixed angle scattering} We start by studying the fixed angle high-energy scattering scenario, in which one considers $s\to-\infty~\text{and}~t\to \infty$, while the $2 \rightarrow 2$ scattering angle $\phi$ is held fixed. In particular, we have\footnote{Here we employ the same parameterisation scheme and notation as in the original paper by Veneziano \cite{Veneziano:1968yb}.}
\begin{align}
    -\frac{t}{s} \sim \sin^2 \frac{\phi}{2} = \frac{1-\cos \phi}{2} = \frac{1-x}{2}~,
\end{align}
where we denote $x = \cos \phi$. With the above constraint, the Mandelstam invariants can be expressed as 
\begin{align}
    s = -E^2~, \qquad t = E^2 \left(\frac{1-x}{2}\right)~,
\end{align}
where $E$ is the centre of mass energy squared. The Veneziano amplitude in this limit takes the form\footnote{In doing so, the Stirling's approximation of the gamma function comes in handy 
\begin{equation*}
    \ln{\Gamma(x)} \sim x\ln{x}+\mathcal{O}(\ln{x})~,~~~\text{as}~x\to \infty~,
\end{equation*}
~~~~~which holds for any $x\in \mathbb{C}/\{\mathbb{Z}_{\leq 0}\}$.
}
\begin{equation}
    \mathcal{A}^{V}_4(E,x) \sim \exp{\left(-E^2 f(x)\right)}~,~~\text{as}~E\to \infty~,
\end{equation}
while we have
\begin{equation}\label{eq:f_high_energy_function}
    f(x) = \left(\frac{1-x}{2}\right)\log\left(\frac{2}{1-x}\right)+ \left(\frac{1+x}{2}\right)\log\left(\frac{2}{1+x}\right)~,
\end{equation}
which is a positive function for physical angles i.e. for $x\in [0,1)$. Plugging in these expressions into the twisted four-point amplitude (\ref{eq:BD_amplitude}) we get the following expression  
\begin{equation}\label{eq:BD_amplitude_high_energy_limit}
    \mathcal{A}^{\text{twisted,AC}}_4(E,x) \sim \exp{\left(-E^2 f(x)\right)}\left(1-h\frac{E^4}{8}(1-x)+\mathcal{O}(h^2)\right)~.
\end{equation}
Clearly, the above amplitude dies off smoothly in the limit $E\to \infty$, thus showcasing exponentially soft UV-finite behaviour, characteristic of tree-level string scattering. However, this happens at a rate $E^4$ slower than the Veneziano amplitude. This differs from the hypergeometric amplitudes stated in \cite{Cheung:2023adk} which showcase power law behaviour for high-energy fixed angle scattering. Moreover, this result is consistent with the asymptotic behaviour of dual resonant amplitudes predicted in \cite{Caron-Huot:2016icg}. 

\paragraph{Regge behaviour} We now consider the Regge limit, which means $s\to -\infty$ while $t$ is held fixed. For the Veneziano amplitude, one obtains in this limit
\begin{equation}
    \mathcal{A}^{\text{V}}_4(s,t) \sim a(t)s^{\frac{t}{2}+1}~,
\end{equation}
where $a(t) = \Gamma\left(-t/2-1\right)$. Therefore, we have for the twisted four-point amplitude
\begin{equation}\label{eq:BD_amplitude_regge_limit}
    \mathcal{A}^{\text{twisted,AC}}_4(s,t) \sim a(t)s^{\frac{t}{2}+1}+h \, b(t)s^{\frac{t}{2}+2}+\mathcal{O}(h^2)~, ~~s\to \infty~~\text{and}~~t=\text{constant}~,
\end{equation}
where we have defined $b(t)\equiv \frac{ta(t)}{4}$. The leading Regge behaviour is dominated by the second term in equation (\ref{eq:BD_amplitude_regge_limit}), which implies that the amplitude has a linear Regge trajectory with shifted intercept, in comparison to its Veneziano counterpart \cite{Veneziano:1968yb}. Moreover, the above trajectory is well behaved for $\text{Re}(t)\leq -4$, which defines the region of physical kinematics for the twisted amplitude. This goes to show that linear Regge behaviour is non-unique to the Veneziano amplitude \cite{pirsa_PIRSA:23070040}.


\section{\texorpdfstring{Coon amplitude from worldsheet $q$-CFTs}{Coon amplitude from worldsheet q-CFTs}}
\label{sec:Coonamp}
After having studied the set of observables arising out of worldsheet $q$-CFTs based on the quantum group $SU(1,1)_q$, we now turn towards asking if any known $q$-deformations of the Veneziano amplitude (or an avatar thereof) arise in such theories. In particular, we focus on the status of the Coon amplitude \cite{Coon:1969yw} -- one of the earliest known one-parameter deformation of the Veneziano amplitude (we refer the reader to appendix \ref{app:coonamplitude} for a quick review on the Coon amplitude) in such $q$-deformed worldsheet theories. To achieve this, we adopt a reverse engineering approach in the sense that we extract the choices for the undetermined constants $\gamma_{i j}$ and the conformally invariant function $F(u,v)$ of the four-point $q$-deformed  ansatz that the integral representation of the Coon amplitude predicts. This allows for a direct comparison to the $q$-deformed observables (and suitable analytic continuations thereof) derived in our analysis so far and forms the basis of this section.

Starting from the integral representation of the Coon amplitude 
\begin{equation}
    \mathcal{A}^{C}_4(s,t) = \int_{0}^{1} d_qz \, \, z^{-\alpha_q(s)-2} (1-qz)_q^{-\alpha_q(t)-2}\label{eq:coon_integral_amp}~,
\end{equation}
we may recast its integrand in terms of the basic hypergeometric functions introduced in section \ref{sec:setup}
\begin{equation}
    \mathcal{A}^{C}_4(s,t) = \int_{0}^{1} d_qz \, \, z^{-\alpha_q(s)-2} {}_{1}\varphi_{0}\left(\alpha_q(t)+2;q^{-\alpha_q(t)-1}z\right)~.
    \label{eq:Coon_with_MS_ansatz}
\end{equation}
In the above, we have simply rewritten the second term of the integrand (\ref{eq:coon_integral_amp}) in terms of the basic hypergemoetric series ${}_{1}\varphi_{0}$ using the identity
\begin{equation}
    {}_{1}\varphi_{0}(a;x) = (1-xq^a)_q^{-a}.
\end{equation}
Now, we consider the general ansatz for a $SU(1,1)$ CFT four point correlator 
\begin{equation}
   \la \phi_1 \dots \phi_4\ra = F(u,v)\prod_{i<j}z_{ij}^{-\gamma_{ij}}
\end{equation}
where $F(u,v)$ is a general function of the $SU(1,1)$ conformal cross-ratios. An appropriate $q$-deformation of the above is 
\begin{equation}
  \la \phi_1 \dots \phi_4\ra_q = F_q(u,v)\prod_{i<j}z_{i}^{-\gamma_{ij}} {}_{1}\varphi_{0} \left(\gamma_{ij};~x_{ij} q^{a_{ij}}\right)~,
  \label{eq:general_ansatz_4pt}
\end{equation}
where, as before, $\gamma_{ij}$ and $a_{ij}$ are the coefficients which we constrain via a direct comparison with the Coon amplitude in (\ref{eq:coon_integral_amp}) and the $q$-Ward identities discussed in section \ref{sec:setup}. We now write our general $q$-string amplitude by utilising the open-string gauge-fixing conditions introduced in section \ref{sec:qCFTDrinfel'd}. Using the $q$-deformed ansatz for the four-point correlator (\ref{eq:general_ansatz_4pt}), we obtain the corresponding gauge-fixed $q$-deformed amplitude
\begin{align}
    \mathcal{A}^{q}_4(s,t) &\sim \int d_q z \lim_{z_1 \to \infty} z_1^2 \la \phi_1 \dots \phi_4\ra_q \nonumber \\
    &= \int d_q z \lim_{z_1 \to \infty} z_1^{2-\sum_{j=2}^4 \gamma_{1 j}} x^{-\gamma_{34}} {}_{1}\varphi_{0} (\gamma_{23}; x q^{a_{23}}) F_q (x)~.
    \label{eq:Coon_with_MS_ansatz_gauge_fixed}
\end{align}
To make contact with the integral representation of the Coon amplitude (\ref{eq:Coon_with_MS_ansatz}), we have two possible sets of choices of the parameters appearing in the above expression for the $q$-deformed amplitude. The first set of choices for the parameters is given by
\begin{align}
    \sum_{j=2}^4 \gamma_{1 j} = 2~,\quad \gamma_{34} = \alpha_q(s)+2~, \quad \gamma_{23} = \alpha_q (t) + 2~,\quad a_{23} = -\alpha_q(t)-1, \quad F_q(x) = 1~.  
    \label{eq:Choice1}
\end{align}
We note that the analysis carried out in section \ref{sec:qCFTDrinfel'd} in our construction of $q$-CFTs is based on the above ansatz where the conformal invariant function is $F_q(x)=1$. This revealed that such an ansatz for the $q$-correlator lead to vanishing $q$-deformed amplitudes. If we decide to move away from the realm of $q$-CFTs, we can compare with the analytically continued twisted amplitude $\mathcal{A}_4^{\text{twisted,AC}}(s,t)$ (\ref{eq:analcontamp}) proposed at the end of section \ref{sec:twsitedamp}. Despite both the twisted $\mathcal{A}_4^{\text{twisted,AC}}(s,t)$ and the Coon amplitude $\mathcal{A}_4^{C}(s,t)$ flowing to the Veneziano amplitude in the $q \to 1$ (or $h \to 0$), the asymptotic behaviours of the amplitudes are drastically different. While the Regge behaviour of the twisted amplitude is linear with shifted intercepts, the Coon amplitude exhibits logarithmic Regge trajectories described by the relation (\ref{eq:reggebehavCoon}). Thus, the above arguments tell us that although the choice of parameters (\ref{eq:Choice1}) is consistent with the Coon amplitude, the corresponding $q$-deformed $SU(1,1)_q$ amplitudes are either vanishing or have different asymptotic properties.

Apart from the above choice of parameters, one could also directly compare with (\ref{eq:Coon_with_MS_ansatz}) and identify the conformally invariant function $F_q(x)$ to be of the form
\begin{align}
    F_q(x) = x^{\gamma_{34}-\alpha_q(s)-2} \frac{{}_{1}\varphi_{0}\left(\alpha_q(t) +2; x q^{-\alpha_q(t) -1}\right)}{{}_{1}\varphi_{0}\left(\gamma_{23};x q^{a_{23}}\right)}~.
    \label{eq:Choice2}
\end{align}
Taking cue of the fact that one recovers usual string theory in the $q \to 1$ limit leads to the fact that $F(x)=1$ following which the parameters above are fixed to be 
\begin{align}
    \gamma_{23} = t+2~,\quad \gamma_{34} = s+2~,\quad a_{23} = -t-1~.
\end{align}
The above choice for the ansatz has been checked to neither satisfy the $q$-Ward identities arising from non-invertible Hopf algebras studied in section \ref{sec:setup} nor the twisted Ward identities arising from invertible Hopf algebras studied in section \ref{sec:qCFTDrinfel'd}. As a result, we are forced to rule out amplitudes arising out of an ansatz where the conformal invariant function takes the form (\ref{eq:Choice2}). 

It is worth reiterating that we have only two possible choices for the ansatz given by (\ref{eq:Choice1}) and (\ref{eq:Choice2}) that is consistent with the Coon amplitude. Working with a four-point ansatz dictated by the choice (\ref{eq:Choice1}) forms the basis of our analysis in sections \ref{sec:setup} and \ref{sec:qCFTDrinfel'd} which reveal that all $q$-deformed amplitudes arising out of $SU(1,1)_q$ worldsheet theories vanish. Even the analytically continued amplitude $\mathcal{A}_4^{\text{twisted,AC}}$ fails to match with the asymptotic properties of the Coon amplitude.
Combined with the fact that the other allowed choice (\ref{eq:Choice2}) gives rise to amplitudes that do not satisfy the Ward identities corresponding to $q$-deformed algebras for any choice of the unknown parameters in the four-point ansatz, we are led to conclude that the Coon amplitude does not belong to the set of observables in $q$-deformed CFTs in the construction considered in our paper.   
\section{Outlook and discussions}
\label{sec:discussions}
In our analysis, we have initiated a systematic study to determine the set of observables that can be constructed out of $q$-deformed worldsheet CFTs with the underlying conformal group being the quantum group $SU(1,1)_q$. Here, we summarize the key results of our paper before discussing some potential avenues for future work.

The analysis begins in section \ref{sec:setup} where we review existing literature on this subject and shed new insights on certain old results. In particular, we look at correlation functions constructed out of Hopf algebra maps that are \textit{non-invertible} when flowing from the classical Lie group to its quantum counterpart $SU(1,1) \to SU(1,1)_q$. Relying on explicit computations for certain well-known deformations and inductive arguments, we provide arguments on the interrelationships in the moduli space of non-invertible Hopf algebras corresponding to Quantum Universal Enveloping (QUE) groups. We conclude that the gauge-fixed, tachyonic four-point $q$-amplitudes constructed out of such non-invertible Hopf algebras vanish due to the reduced constraint $s+t=-4$.

We go on to analyse, for the first time (as far as we know), the $q$-deformed correlation functions and $q$-amplitudes constructed out of \textit{invertible} Hopf algebras called Drinfel'd twists \cite{drinfel'd1989quasi}. To our amusement, we find that although such twisted $q$-amplitudes flow to the Veneziano amplitude when $q=1$, they identically \textit{vanish} for $q \in [0,1)$, again on the support of the constraint $s+t=-4$. Combined with the analogous result for non-invertible Hopf algebras, this leads us to conclude that tachyonic $S$-matrices for $q$-deformed CFTs with an underlying conformal group $SU(1,1)_q$ is \textit{trivial}, i.e., it is simply the identity matrix.

While the twisted amplitude arising out of the coproduct map given by the Drinfel'd twist vanishes, we however postulate a well-behaved class of amplitudes (\ref{eq:analcontampgen}) on analytically continuing to the full $(s,t,u)$ kinematic phase space. We focus our attention to a more restricted space of $q$-deformed twisted amplitudes given by
    \begin{align}
    \mathcal{A}^{\text{twisted,AC}}_4(s,t) = \mathcal{A}^{V}_4(s,t) \left[1+ h \left(\frac{st}{4} \right)+\mathcal{O}(h^2)\right]~. 
\end{align}
    This four-point tachyonic amplitude serves as a $\mathcal{O}(1-q)$ correction to the Veneziano amplitude $\mathcal{A}^{V}_4(s,t)$ and has well-defined high-energy and Regge behaviour, as discussed in section \ref{sec:asympproptwistedamp}.

We conclude our analysis in section \ref{sec:Coonamp} by arguing against the possibility that the Coon amplitude, a well-studied one-parameter deformation of the Veneziano amplitude, can arise out of worldsheet $q$-CFTs whose global conformal group is $SU(1,1)_q$. We also compare the Coon amplitude with the new twisted amplitudes $\mathcal{A}^{\text{twisted,AC}}_4(s,t)$ proposed in our work and find that they have varying asymptotic behaviour. With the above summary of results, we now turn towards discussing some potential directions for future work.

Our analysis in sections \ref{sec:setup} and \ref{sec:qCFTDrinfel'd} considers the $S$-matrix of $q$-string theory corresponding to the scalar primary sector of the string multiplet. It is then natural to investigate the $S$-matrix for the higher spin modes, which would come out of studying the worldsheet $q$-CFT for a spin $J$ conformal primary $\phi_{\Delta}^J(z)$. The transformation property under the action of a $q$-Virasoro generator $\mathcal{L}_n$ in DJ deformation for $\phi^{J}_{\Delta}$ is given as \cite{Oh:1992hn}
    \begin{equation}
        \mathcal{L}_n\cdot\phi^{J}_{\Delta}(z) = z^n[z\partial_z+(\Delta+J)(n+1)]_q\phi^{J}_{\Delta}(z)~.
    \end{equation}
One can then construct the conformal $q$-Ward identities using the Hopf algebra (\ref{eq:hopf_algebra_of_DJ}), as outlined in our analysis. If one finds trivial $S$-matrices for the spin $J$ sector, this would conclude that any $S$-matrix derived from the $q$-string theory arising from the $q$-deformation of the worldsheet conformal group is trivial, thus extending the arguments of this paper beyond the tachyonic sector. However, a non-trivial result, would mean the tachyon entirely decouples in this theory. This could lead to a possible resolution for the tachyon problem that bosonic string theory suffers from \cite{polchinski_1998}, without having to resort to introducing the supersymmetric multiplet \cite{polchinski_1998_2}.

Another important future direction is the further analysis of the new class of dual resonant twisted four-point amplitudes $\mathcal{A}_{4}^{\text{twisted,AC}}$ given by (\ref{eq:analcontampgen}) from a bootstrap perspective. In section \ref{sec:asympproptwistedamp}, we have mainly focused on the specific case $\alpha,\beta=0$ and studied its asymptotic behaviour. One can further try studying the more general class of amplitudes for $\alpha,\beta \neq 0$. The functions $f$ and $g$ are polynomials and take the form
    \begin{equation}
        f(x) = \sum_{n=0}^{N_{max}}a_n\,x^n~,~~~~g(x) = \sum_{m=0}^{M_{max}}b_m\,x^m~,
    \end{equation}
for $N_{max},M_{max}\in \mathbb{N}$ and $x=s+t$.\footnote{Here, taking $N_{max},M_{max}\to \infty$ would lead to the most general ansatz for this amplitude, but for practicality it is always beneficial to analyse the finite order case as an approximation; for more details please refer to \cite{Atkinson:1970zza,Paulos:2017fhb}.} Then using unitarity, analyticity and polynomial boundedness one can constraint the coefficients $(a_n,b_m)$ along the lines of \cite{Atkinson:1970zza}. It would then be interesting to ask, if there exists a field theoretic or worldsheet-like derivation for the resulting amplitude and how such amplitudes connect to the analysis presented in \cite{pirsa_PIRSA:23070040}.
    
Another future line of analysis is the computation of higher-point correlators in $q$-CFTs constructed in our paper. One could then go ahead and explicitly check, if the corresponding higher-point amplitudes, like their lower point cousins, vanish on the support of the higher point analog of the reduced constraint. However, we know from the factorisation property of higher point amplitudes in QFT, it is highly likely that if the lower-point amplitudes vanish, so will their higher counterparts. However, an explicit proof of the vanishing of $n$-point $S$-matrix for $n\geq 5$ is still lacking in our analysis.
    
    
\section*{Acknowledgements} 
The authors would like to thank Adam Ball, Clifford Cheung, Juan Maldacena, Grant Remmen, and Alexander Zhiboedov for useful comments and discussions. The authors are especially grateful to Nick Geiser, Sanjaye Ramgoolam, Marcus Spradlin, Akshay Yelleshpur Srikant and Anastasia Volovich for valuable feedback and comments on the manuscript. Finally, the authors are indebted to Sanjaye Ramgoolam for introducing them to the Drinfel'd twist and related references, which inspired Section \ref{sec:qCFTDrinfel'd} of this paper. This work was supported by the US Department of Energy under contract {DE}-{SC}0010010 Task F.

\appendix
\section{Quantum groups}\label{app:quantum_group}

In this appendix, we provide a general overview of the mathematical details associated to quantum groups, in particular introducing the formal definitions of a bi-algebra and Hopf algebras. We elaborate on the existence of the $\mathcal{R}$-matrix and the notion of quasi-triangularity, before establishing the notion of (in)equivalent deformations. While most of the important machinery and formulae needed for actual computations have been presented in sections \ref{sec:setup} and \ref{sec:qCFTDrinfel'd}, this appendix is largely presented for the sake of completeness and to facilitate complementary reading with the above sections. 

In classical physics, the observables are often defined as operators $\mathcal{O}$ on a commutative ring $R$ such that:
\begin{align}
    [\mathcal{O}_1,\mathcal{O}_2](p)\equiv \mathcal{O}_1(p)\mathcal{O}_2(p)-\mathcal{O}_2(p)\mathcal{O}_1(p)=0~,~~\forall~\mathcal{O}_1,\mathcal{O}_2\in R,~\forall~p\in \mathcal{M}~,
\end{align}
where $\mathcal{M}$ is the background manifold on which the theory is defined. However, in quantum physics $\mathcal{O}$ is defined on an associative non-commutative ring $R_{h}$ composed of hermitian operators\footnote{This is often the Hilbert space of observables $\mathscr{H}$.} such that the above commutator is non-vanishing for any generic $p$. Mathematically this process of moving from classical to quantum analogs is described as the process of deformation \cite{Rakic:2001zz}.

Consider the classical $SO(3)$ Lie algebra whose orbit describes the classical (commutative) 2-sphere $S^2$:
\begin{equation}
    [L_0,L_+] = L_+~,~~~[L_+,L_-] = L_0~,~~~[L_-,L_0] = L_-~,\label{eq:classical_so3_algebra}
\end{equation}
and the Casimir invariant is 
\begin{equation}
    C(L) \equiv L_x^2+L_y^2+L_z^2 = L_+L_-+L_-L_++L_0^2 \label{eq:classical_so3_casimir}~.
\end{equation}
In order to deform the isotropy of this $SO(3)$ symmetric expression, one must introduce an ``anisotropic" term, so a possible deformation is
\begin{equation}
    C_q(L) = L_+L_-+L_-L_++\frac{q+q^{-1}}{2}\left(\frac{q^{L_0}-q^{-L_0}}{q-q^{-1}}\right)^2~,
    \label{eq:quantum_so3_casimir}
\end{equation}
where we have introduced $q=e^h$ which parametrises the amount of anisotropy. Clearly in the $h\to 0$ limit, one recovers the classical Casimir (\ref{eq:classical_so3_casimir}). This amounts to the following deformation of the Lie algebra 
\begin{equation}
    [\mathcal{L}_{0},\mathcal{L}_{+}] = \mathcal{L}_{+}~,~~~[\mathcal{L}_{+},\mathcal{L}_{-}] = \frac{1}{2}[2\mathcal{L}_{0}]_q~,~~~[\mathcal{L}_{-},\mathcal{L}_{0}] = \mathcal{L}_{-}~,\label{eq:quantum_so3_algebra}
\end{equation}
where $\{\mathcal{L}_{0},\mathcal{L}_{+},\mathcal{L}_{-}\}$ are the new generators. This was first introduced by Drinfel'd \cite{drinfel'd1985hopf} and later independently by Jimbo \cite{Jimbo:1985zk}, and hence goes by the name Drinfel'd-Jimbo type deformation of $SU(1,1)$ Lie algebra (the universal enveloping algebra of $SO(3)$). More generally, it is a (quasi-triangular) Hopf algebra, which we now define in the subsequent paragraphs. 

Let $R$\footnote{Classically, this ring is mostly the reals $\mathbb{R}$ or the complex numbers $\mathbb{C}$.} be a commutative ring with unit $1$. Let there be an associative algebra $A$ over $R$ with unit $e$ known as the $R$-module. Further, we define the notion of multiplication on $A$ as the associative map $\cdot:A\otimes A\to A$ and unit as $e: R\to A$. One can further define the inverses of these maps namely the comultiplication $\Delta: A \to A\otimes A$, and counit $\Bar{e}:A\to R$. Likewise $\Delta$ is coassociative i.e. $(\text{id}\otimes\Delta)\Delta(a)= (\Delta\otimes\text{id})\Delta(a)$ $\forall$ $a\in A$. These two objects together define a coalgebra of their own. An algebra defined this way which is both a coalgebra and an algebra at the same time is known as a \textit{bi-algebra}. The general form of a coproduct is 
\begin{equation}
    \Delta(a) = \sum_{i}a_i^{(1)}\otimes a_i^{(2)}~,
\end{equation}
where $a^{(1)}_i$ is an element of $A$ in the first slot of $A\otimes A$ and likewise with $a^{(2)}_i$. In the notation due to Swedler \cite{swedler1969hopf}, one suppresses the dummy index $i$ and the summation symbol:
\begin{equation}
    \Delta(a) = a^{(1)}\otimes a^{(2)}~.
\end{equation}
Likewise, using the above notation, one can define the notion of cocommutativity
\begin{equation}
    a^{(1)}\otimes a^{(2)} = a^{(2)}\otimes a^{(1)}~,
\end{equation}
for all $a^{(1,2)}\in A$, and one can restate the above more compactly via the swap map $\tau:a^{(1)}\otimes a^{(2)} \mapsto a^{(2)}\otimes a^{(1)}$ as $\tau\Delta = \Delta$.
Another object necessary for defining a Hopf algebra is the antipode map $S$, such that $(\text{id}\otimes S)\Delta(a)= (S\otimes \text{id})\Delta(a)$, and is defined by the following property:
\begin{equation}
    a^{(1)}\cdot S(a^{(2)}) = S(a^{(1)})\cdot a^{(2)} = e\cdot \Bar{e}(a)~.\label{eq:antipode_definition}
\end{equation}
A \textit{Hopf algebra} then is a bi-algebra  $(A, e, \Bar{e}, \Delta)$ endowed with the antipode map $S$ defined by (\ref{eq:antipode_definition}). In general, a Lie algebra $\mathfrak{g}$ is not a Hopf algebra; however, it turns out that any universal enveloping algebra $\mathcal{U}(\mathfrak{g})$ thereof,  without its ideal $I(\mathfrak{g})$\footnote{An ideal $I$ of a given Lie algebra $\mathfrak{g}$ is a sub-algebra such that $\forall X\in \mathfrak{g}$, $\exists Y\in I$ so that $[X,Y] \in I$.}, is a non-commutative co-commutative Hopf algebra, given as follows
\begin{align}
\Bar{e}(e)&=1~,\qquad\qquad\bar{e}(X)=0~,\nonumber\\
    \Delta(e) &= e\otimes e~,\qquad\Delta(X)=X\otimes e+e\otimes X~,\nonumber\\
    S(e) &= e~,\qquad\qquad S(X)=-X~,
\end{align}
for any $X \in \mathfrak{g}$. A $q$-deformation of the above Hopf algebra, for $\mathfrak{g}=\mathfrak{su}(2)$, is given as
\begin{align}
\bar{e}(e)&=1~,\qquad\qquad\qquad~~~~~~\Bar{e}(\mathcal{L}_{0})=\Bar{e}(\mathcal{L}_{\pm})=0~,\nonumber\\
    \Delta_q(\mathcal{L}_{0}) &= \mathcal{L}_{0}\otimes e+e\otimes \mathcal{L}_{0}~,\qquad \Delta_q(\mathcal{L}_{\pm}) = \mathcal{L}_{\pm}\otimes q^{\mathcal{L}_{0}/2}+q^{-\mathcal{L}_{0}/2}\otimes \mathcal{L}_{\pm}~,\nonumber\\
    S(\mathcal{L}_{0})&=-\mathcal{L}_{0}~,\qquad\qquad\qquad~~ S(\mathcal{L}_{\pm}) = -q^{\pm 1}\mathcal{L}_{\pm}~.\label{eq:su2q_hopf_algebra}
\end{align}
Of course, as stated before, this is the Hopf algebra corresponding to the universal enveloping group of $\mathfrak{su}(2)_q$ i.e. $\mathcal{U}(\mathfrak{su}(2))_q$ and not $\mathfrak{su}(2)_q$ itself. The Hopf algebra (\ref{eq:su2q_hopf_algebra}) along with the group algebra (\ref{eq:quantum_so3_algebra}) define the \textit{quantum group $\mathfrak{su}(2)_q$}. Moreover, the coproduct for the above map is an algebra homomorphism, i.e.  it preserves the group algebra (\ref{eq:quantum_so3_algebra})
\begin{equation}
    [\Delta_q(\mathcal{L}_{0}),\Delta_q(\mathcal{L}_{\pm})] = \pm\Delta_q(\mathcal{L}_{\pm})~,~~~[\Delta_q(\mathcal{L}_{+}),\Delta_q(\mathcal{L}_{-})] = \frac{1}{2}[2\Delta_q(\mathcal{L}_0)]_q~,
\end{equation}
and similarly for the antipode map $S$. We note that the algebra (\ref{eq:quantum_so3_algebra}) is invariant under the operation $q\leftrightarrow q^{-1}$ but the Hopf algebra is not; it turns out that the two Hopf algebras are related via the universal $\mathcal{R}$-matrix:
\begin{equation}
    \Delta_{q^{-1}}(X) = \mathcal{R}\Delta_q(X)\mathcal{R}^{-1}~,
\end{equation}
or equivalently 
\begin{equation}
    \tau\cdot\Delta_q(X) = \mathcal{R}\Delta_q(X)\mathcal{R}^{-1}~,\label{eq:$R$-matrix_definition}
\end{equation}
where $\mathcal{R} \in A\otimes A$. A Hopf algebra which allows for the existence of such an $\mathcal{R}$-matrix is called a \textit{quasi-triangular} Hopf algebra. For the quantum group $\mathfrak{su}(2)_q$ we have for the universal $\mathcal{R}$-matrix:
\begin{equation}
    \mathcal{R} = q^{\frac{j_0\otimes j_0}{2}}\sum_{n=0}^{\infty}\frac{(1-q^2)^n}{[n]_q!}q^{n(n-1)/2}(q^{j_{0}}j_{+}\otimes q^{-j_{0}}j_{-})^n~.
\end{equation}
In the matrix representation, one can use the fundamental representation of the generators $\bigg\{j_{0} = \begin{pmatrix}
1 & 0 \\
0 & -1 
\end{pmatrix}, j_{+}= \begin{pmatrix}
0 & 1 \\
0 & 0 
\end{pmatrix}, j_{-}= \begin{pmatrix}
0 & 0 \\
1 & 0 
\end{pmatrix}\bigg\}$ to get the following form for the $\mathcal{R}$-matrix: 
\begin{equation}
  \mathcal{R} =  \frac{1}{\sqrt{q}}\begin{pmatrix}
q & 0 & 0 & 0\\
0 & 1 & q-q^{-1} & 0\\
0 & 0 & 1 & 0\\
0 & 0 & 0 & q
\end{pmatrix}~.
\end{equation}
Knowing that $\mathcal{R}$ is an element in $A\otimes A$, we can also consider a basis decomposition such that for a given basis $\{e_i\}$ of $A$ we have 
 \begin{equation}
     \mathcal{R} = \sum_{i,j}R^{(ij)}e_i\otimes e_j ~~~\text{where}~~~R^{(ij)} \in \mathbb{C}~.
 \end{equation}
Furthermore, the $\mathcal{R}$-matrix also satisfies the relations
\begin{equation}
    (\Delta_q\otimes \text{id})(\mathcal{R}) =\mathcal{R}_{12}\mathcal{R}_{23}~,~~~~ (\text{id}\otimes \Delta_q)(\mathcal{R}) =\mathcal{R}_{13}\mathcal{R}_{12}~\label{eq: quasi_triangularity_equations-1},
\end{equation}
where
\begin{equation}
    \mathcal{R}_{12} = \sum_{i,j}R^{(ij)}e_i\otimes e_j\otimes e~,~~\mathcal{R}_{13} = \sum_{i,j}R^{(ij)}e_i\otimes e\otimes e_j~,~~\mathcal{R}_{23} = \sum_{i,j}R^{(ij)}e\otimes e_i\otimes e_j~,
\end{equation}
and $e$ is the identity element in $A$ as defined before. Then using (\ref{eq: quasi_triangularity_equations-1}) together with (\ref{eq:$R$-matrix_definition}), one can derive the well-known Yang-Baxter equations \cite{Jimbo:1985zk}
\begin{equation}
\mathcal{R}_{12}\mathcal{R}_{13}\mathcal{R}_{23} = \mathcal{R}_{23}\mathcal{R}_{13}\mathcal{R}_{12}~.
\end{equation}
This is often seen as a test of quasi-triangularity of a given quantum Hopf algebra along with the condition
\begin{equation}
 \mathcal{R}_{21} = \tau \mathcal{R}_{12} = \mathcal{R}_{12}^{-1}~.
\end{equation}
The two Hopf algebras $\Delta_q$ and $\Delta_{q^{-1}}$ are equivalent in this sense, i.e. they are related by a universal $\mathcal{R}$-matrix. 

More generally, at the level of Lie algebras, there exists an invertible map between the usual $\mathfrak{g}$ and its various $q$-deformed counterparts given by the deforming functional $Q:\mathfrak{g}\to \mathfrak{g}_q$. More explicitly, the map at the level of coproducts works as
\begin{equation}
    Q(\Delta(\mathfrak{g})) = Q(e\otimes Q^{-1}(\mathfrak{g}_q)+Q^{-1}(\mathfrak{g}_q)\otimes e)~,
\end{equation}
it is often cumbersome to construct such invertible mappings at the level of Hopf algebras and in some cases even impossible\footnote{Specifically for the case when $q=e^{2\pi i/n}$ for $n\in \mathbb{N}$ as discussed in section \ref{sec:spaceofQUE-algebras}.}\cite{Zachos:1990wi,Curtright:1989sw}. However, specifically for the case of $\mathfrak{su}(2)$, the above is simplified via the similarity transform
\begin{equation}
    Q(\Delta(\mathfrak{g})) = U_q^{-1}\Delta_q(\mathfrak{g}_q) \, U_q~,
    \label{eq:similarity_transform}
\end{equation}
where $U_q \in \text{End}(\mathfrak{g}_q\times \mathfrak{g}_q,\mathbb{C})$ is a unitary operator that maps the Clebsch-Gordon operator of $\mathfrak{su}(2)$ to that of $\mathfrak{su}(2)_q$. $\Delta(\mathfrak{g}_q)$ is often chosen to be a given base deformation of the $\mathfrak{su}(2)$ algebra for which the Hopf-algebra is well known. In the literature, this is often chosen to be the Drinfel'd-Jimbo type deformation \cite{Zachos:1990wi}. The deformations whose Hopf algebra is related to the base case algebra via (\ref{eq:similarity_transform}) are said to be \textit{equivalent} \cite{Zachos:1990wi,Dlamini:2019zuk}. There are, however, deformations of the Hopf algebra for which there does not exist such a similarity transformation --- these are the \textit{inequivalent} deformations and are related to the classical Hopf algebra via the so-called \textit{Drinfel'd twists}, which is the topic of the next appendix. These definitions are useful for the discussion in section \ref{sec:spaceofQUE-algebras}. 

\section{Drinfel'd twists and invertible Hopf algebras}\label{app:Drinfel'd_twist}
In this appendix, we attempt to provide a formal definition of the Drinfel'd twist and an explicit perturbative construction of the associated coproduct. While the essential formulae needed for the construction of Ward identities using the Drinfel'd twists have been presented in section \ref{sec:qCFTDrinfel'd}, this appendix serves as a mathematical primer on the subject. 

It was shown by Drinfel'd, in his seminal work \cite{drinfel'd1985hopf,drinfel'd1989quasi,drinfeld1990quasi} using quantum cohomology and WZW theory, that there exists an isomorphism between the quantum symmetries (quasi-triangular Hopf algebra) $\mathfrak{g}_q$ and the corresponding universal covering group of any given semi-simple Lie algebra $\mathfrak{g}$.\footnote{A semi-simple Lie algebra is any Lie algebra, which does not have any non-trivial ideals, with the trivial ones corresponding to $\mathfrak{g}$ and $\{0\}$.} The two are related by a pair $(Q, \mathcal{F})$, where $Q$ is the isomorphism between the quantum and the classical group introduced in appendix \ref{app:quantum_group}, and $\mathcal{F}$ is the so-called Drinfel'd twist which relates the two Hopf algebras. However, the problem of finding such a pair is rather cumbersome and is not solved in its full generality. One therefore relies highly on taking a perturbative approach, which we review below. 

Let us consider how the in-equivalent quasi-triangular Hopf algebras mentioned in the previous appendix are related to one another. Consider $A$ to be a quasi-triangular Hopf algebra with the $\mathcal{R}$-matrix. Then it can be shown that $\exists$ an invertible bi-linear map $\mathcal{F}\in A\otimes A$ that satisfies $(S\otimes \text{id})\mathcal{F}= (\text{id}\otimes S)\mathcal{F}=\mathbb{I}$ known as the Drinfel'd twist. We can now use this twist in order to deform the coproduct as \cite{Dlamini:2019zuk}
\begin{equation}
    \tilde{\Delta}(Q(X)) = \mathcal{F}\Delta(X)\mathcal{F}^{-1}~,~~~\forall X\in \mathfrak{g}~, \label{eq:twist_definition}
\end{equation}
known as the twisted coproduct, which is related to the classical coproduct via an invertible map. The $\mathcal{R}$-matrix may also be twisted and written as
\begin{equation}
    \Tilde{\mathcal{R}}_{12} = \mathcal{F}_{21}\mathcal{R}_{12}\mathcal{F}_{12}^{-1}~.
\end{equation}
It can be easily checked that this twisting operation preserves quasi-triangularity 
\begin{equation}
    \Tilde{\mathcal{R}}_{21} \equiv \tau  \mathcal{F}_{21}\mathcal{R}_{12}\mathcal{F}_{12}^{-1} =  \mathcal{F}_{12}\mathcal{R}_{21}\mathcal{F}_{21}^{-1} = \mathcal{F}_{12}\mathcal{R}_{12}^{-1}\mathcal{F}_{21}^{-1} = \Tilde{\mathcal{R}}_{12}^{-1}~,
\end{equation}
where in the final two steps we have used triangularity on the classical $\mathcal{R}$-matrix. To find the general pairing of the invertible isomorphism $Q$ defined in (\ref{eq:similarity_transform}), and the twist $\mathcal{F}$, one works perturbatively in the parameter $h$. It was already shown by Drinfel'd in \cite{drinfel'd1989quasi}, that to first order in $h$, the twist is given as 
\begin{equation}
    \mathcal{F} = \mathbb{I}+h \mathcal{R}~,~~~\text{for}~~~\mathcal{R} = L_+\otimes L_--L_-\otimes L_+~.\label{eq:first_order_twist}
\end{equation}
More generally, up to second order in $h$ one has
\begin{equation}
    \mathcal{F} = \mathcal{F}_0+h\mathcal{F}_1+h^2\mathcal{F}_2+\mathcal{O}(h^3)~.
\end{equation}
Using (\ref{eq:twist_definition}), the first two terms in order $h$ are constrained by the following set of equations 
\begin{align}
    [\mathcal{F}_0,\Delta(X)] &= 0~,~~~\forall X\in \mathfrak{su}(2)~,\\
    [\mathcal{F}_1,\Delta(L_0)]&=0~,\\
    [\mathcal{F}_1,\Delta(L_{\pm})] &= (L_{\pm}\otimes L_{0}-L_0\otimes L_{\pm})\mathcal{F}_0~,
\end{align}
and one can easily check that the minimal solution to the above constraints is provided by (\ref{eq:first_order_twist}). It is this solution that we employ in constructing twisted Ward identities in section \ref{sec:qCFTDrinfel'd}. The recursive constraints on the second term are too complicated to be mentioned here; they were first presented in \cite{1997grgp.conf..293D} and the solution is given as 
\begin{align}
    \mathcal{F}_2 &= \frac{1}{2}(\mathbb{I}\otimes L_0^2+L_0^2\otimes \mathbb{I})+\frac{1}{3}(L_+\otimes L_0 L_--L_0 L_+\otimes L_-+L_0 L_-\otimes L_+-L_-\otimes L_0 L_+)\nonumber\\
    &~~~+\frac{1}{6}L_0 \otimes L_0(1-3K)-\frac{11}{24}K+\frac{1}{2}\big((1+K)^2-1-2\mathbb{I}\otimes \mathbb{I}\big)~,
\end{align}
where 
\begin{equation}
    K = 2(L_+\otimes L_-+L_-\otimes L_++L_0\otimes L_0)~,
\end{equation}
is the Cartan-Killing metric. 

\section{A primer on the Coon amplitude}
\label{app:coonamplitude}
The Coon amplitude \cite{Coon:1969yw} is a particular $q$-deformation of the well-known Veneziano amplitude, which is known to describe four-point scattering of strings. It admits the following infinite product representation 
\begin{equation}
    \mathcal{A}^{C}_4(s,t) = (1-q)\exp\left(\frac{\log{\sigma}\log{\tau}}{\log{q}}\right)\prod_{n=0}^{\infty}\frac{\left(1-\frac{q^n}{\sigma\tau}\right)(1-q^{n+1})}{\left(1-\frac{q^n}{\sigma}\right)\left(1-\frac{q^n}{\tau}\right)}~,
    \label{coonkakhoon}
\end{equation}
where we have 
\begin{equation}\label{eq:linear_regge_trajectories}
    \sigma = 1-(1-q)(\alpha_0+s)~,\quad\quad\tau = 1-(1-q)(\alpha_0+t)~~~,
\end{equation}
and $q \in \mathbb{C}$ is an arbitrary parameter. For the sake of unitarity, one must restrict $\text{Re}(q)\in [0,1]$ \cite{Bhardwaj:2022lbz,Figueroa:2022onw,Jepsen:2023sia,Chakravarty:2022vrp}. The Coon amplitude smoothly interpolates between the Veneziano amplitude as $q\to 1$ and scalar field theory as $q\to 0$:
\begin{equation}
    \mathcal{A}^{C}_4(s,t) \overset{q\to0}{\longrightarrow}\frac{1}{s+\alpha_0}+\frac{1}{t+\alpha_0}+1~.
\end{equation}
Utilizing the infinite product representation of the $q$-deformed gamma function one may rewrite (\ref{coonkakhoon}) as \cite{Geiser:2022icl}
\begin{align}
\mathcal{A}_4^\textrm{C}(s,t) &= q^{\alpha_q(s) \alpha_q(t) - \alpha_q(s) - \alpha_q(t)} \frac{\Gamma_q(-\alpha_q(s)) \Gamma_q(-\alpha_q(t))}{\Gamma_q(1-\alpha_q(s)-\alpha_q(t))}~. 
\label{coonq-gamma} 
\end{align}
This can be further recast as a $q$-beta function.  Utilizing its integral representation one can rewrite the amplitude as a ``Koba-Nielsen'' like integral \cite{Romans:1988qs}:
\begin{align}
\mathcal{A}^\textrm{C}(s,t) = \frac{q^{\alpha_q(s) \alpha_q(t) - \alpha_q(s) - \alpha_q(t)}}{[-\alpha_q(s)]_q} \int_{0}^{1} d_qz \, \, z^{-\alpha_q(s)} (1-qz)_q^{-\alpha_q(t)-1}~.
\label{coonintegralrep}
\end{align}
In the representation above, it is clear that the Coon amplitude has non-linear (logarithmic) Regge trajectories: 
\begin{equation}
    \alpha_q(s) \equiv \frac{\ln{\sigma}}{\ln{q}} = \frac{\ln{(1-(1-q)(\alpha_0+s))}}{\ln{q}}~,~\alpha_q(t) \equiv \frac{\ln{\tau}}{\ln{q}} = \frac{\ln{(1-(1-q)(\alpha_0+t))}}{\ln{q}}~,
\end{equation}
in contrast to the linear Regge trajectories $s=\lim_{q \to 1} \alpha_q(s)$ of the Veneziano amplitude (\ref{eq:linear_regge_trajectories}). Unlike the Veneziano amplitude, which has poles at integer values for the Mandelstam variables $s$ and $t$, the Coon amplitude has poles at $q$-integers that were defined in section \ref{sec: DJ_deformation}. The Regge behavior of the amplitude for large $s$ is given as 
\begin{equation}
    \mathcal{A}^\textrm{C}(s,t) \sim a(t)s^{\alpha_q(t)}~,~~~\text{as}~s\to -\infty~~\text{and}~~t=\text{constant}~,
    \label{eq:reggebehavCoon}
\end{equation}
where $a(t)$ is some function of $t$ \cite{Coon:1969yw}.   

\bibliography{main}{}
\bibliographystyle{utphys}

\end{document}